\documentclass[journal]{IEEEtran}
\usepackage{amsmath}
\usepackage{amssymb}
\usepackage{amsfonts}
\usepackage{amsthm}
\usepackage{algorithm}
\usepackage{algorithmic}
\usepackage{graphicx}
\usepackage{array}

\usepackage[shortlabels]{enumitem}
\usepackage{cite}
\usepackage{color}
\usepackage{epsfig,epstopdf}
\usepackage{url}
\usepackage{subfigure}
\usepackage{balance}
\usepackage{multirow}
\usepackage{verbatim}

\IEEEoverridecommandlockouts

\begin{document}

\title{Analysis and Visualization of Deep Neural Networks in Device-Free Wi-Fi Indoor Localization}

\author{Shing-Jiuan Liu, Ronald Y. Chang,~\IEEEmembership{Member,~IEEE}, and Feng-Tsun Chien,~\IEEEmembership{Member,~IEEE}
\thanks{This work was supported in part by the Center for mmWave Smart Radar Systems and Technologies, under the Featured Areas Research Center Program within the framework of the Higher Education Sprout Project by the Ministry of Education (MOE), Taiwan, and in part by the Ministry of Science and Technology (MOST), Taiwan, under Grants MOST 106-2628-E-001-001-MY3, MOST 107-3017-F-009-001, and MOST 107-2221-E-009-037-MY2.}
\thanks{S.-J. Liu and R. Y. Chang are with the Research Center for Information Technology Innovation, Academia Sinica, Taipei, Taiwan (e-mail: \{cindy1445858, rchang\}@citi.sinica.edu.tw).}
\thanks{F.-T. Chien is with the Institute of Electronics, National Chiao Tung University, Hsinchu, Taiwan (e-mail: ftchien@mail.nctu.edu.tw).}}

\maketitle

\begin{abstract}
Device-free Wi-Fi indoor localization has received significant attention as a key enabling technology for many Internet of Things (IoT) applications. Machine learning-based location estimators, such as the deep neural network (DNN), carry proven potential in achieving high-precision localization performance by automatically learning discriminative features from the noisy wireless signal measurements. However, the inner workings of DNNs are not transparent and not adequately understood especially in the indoor localization application. In this paper, we provide quantitative and visual explanations for the DNN learning process as well as the critical features that DNN has learned during the process. Toward this end, we propose to use several visualization techniques, including: 1) dimensionality reduction visualization, to project the high-dimensional feature space to the 2D space to facilitate visualization and interpretation, and 2) visual analytics and information visualization, to quantify relative contributions of each feature with the proposed feature manipulation procedures. The results provide insightful views and plausible explanations of the DNN in device-free Wi-Fi indoor localization using channel state information (CSI) fingerprints.
\end{abstract}


\IEEEpeerreviewmaketitle

\section{Introduction}

Indoor localization is an essential component of many Internet of Things (IoT) applications \cite{macagnano2014IOT}. Applications such as healthcare management \cite{Wyffels2014health}, object tracking \cite{Zheng2017track}, and IoT-based smart environment \cite{Alletto2016smartplace} all require precise indoor location information. From the device perspective, wireless indoor localization can be classified as either device-based or device-free, depending on whether or not a tracking device needs to be attached to the target to be localized \cite{Xiao2016devicefree}. Device-based methods generally have the advantages of higher accuracy and robustness against environmental interferences and dynamics as compared to device-free methods. However, device-free methods have the advantages of less hardware costs, less power demands, better privacy, and provision of real-time positioning and tracking, and find a wide array of IoT applications such as intrusion detection, elderly care, etc. 

From the mathematical techniques perspective, wireless indoor localization can be classified as using triangulation/trilateration or fingerprinting for position calculation. The triangulation/trilateration technique calculates the location of a target based on distances and/or angles of the target with respect to signal sources. The fingerprinting technique matches the online testing measurements with the database created from offline site survey to determine the location of a target. For fingerprinting indoor localization, machine learning-based approaches have been studied \cite{Mager2015ml,Wur2018ml,Zhou2017ml,Saha2003DNNlocal,Zhang2016,Wang2017DNNlocal,Chang2018,Wang2017DNNfree,Gao2017DNNfree}. In \cite{Mager2015ml}, machine learning classifiers such as $k$-nearest neighbor ($k$-NN), support vector machine (SVM), linear discriminant analysis (LDA), and random forests for localization in changing environments were studied and compared. Wu {\it et al.} \cite{Wur2018ml} proposed an indoor localization system using the Naive Bayes classifier and compared it with $k$-NN and SVM. Zhou {\it et al.} \cite{Zhou2017ml} applied SVM to device-free localization based on channel state information (CSI) fingerprinting. In \cite{Saha2003DNNlocal,Zhang2016,Wang2017DNNlocal,Chang2018,Wang2017DNNfree,Gao2017DNNfree}, deep neural network (DNN)-based approaches to indoor localization have been proposed. Saha {\it et al.} \cite{Saha2003DNNlocal} proposed a localization scheme based on the received signal strength (RSS) of Wi-Fi using a neural network-based classifier, which was shown to outperform the nearest neighbor classifier and the histogram matching method. Zhang {\it et al.} \cite{Zhang2016} developed an RSS-based Wi-Fi localization scheme that combines DNN and hidden Markov model (HMM). It was shown that DNN outperforms other machine learning approaches including $k$-NN, SVM, and locally linear embedding (LLE) each in combination with HMM, in terms of root mean square error (RMSE). Wang {\it et al.} \cite{Wang2017DNNlocal} proposed a novel deep learning-based fingerprinting indoor localization system based on the CSI of Wi-Fi. A multilayer framework of a deep network was developed for learning discriminative features so as to effectively reduce the distance error in comparison with the probabilistic methods. Chang {\it et al.} \cite{Chang2018} proposed a device-free DNN-based Wi-Fi fingerprinting localization system. CSI pre-processing and data augmentation (including noise injection and interperson interpolation) were incorporated into the DNN framework for enhanced robustness and performance. Wang {\it et al.} \cite{Wang2017DNNfree} designed a machine learning framework for simultaneous location, activity, and gesture recognition by integrating a sparse autoencoder network with the softmax-regression-based classification module. The learning process can automatically learn the discriminative features from the RSS measurements and achieve an accuracy higher than $85\%$, which is better than the system without utilizing the learning process. Gao {\it et al.} \cite{Gao2017DNNfree} used a multilayer deep learning-based image processing framework to learn the optimized deep features from the radio images (transformed from the amplitude and phase information of the measured CSI) to estimate the location and activity of a target person. Note that \cite{Saha2003DNNlocal,Zhang2016,Wang2017DNNlocal} are device-based and \cite{Mager2015ml,Wur2018ml,Zhou2017ml,Chang2018,Wang2017DNNfree,Gao2017DNNfree} are device-free indoor localization schemes.

DNNs have achieved extraordinary results in many areas. While the DNN can perform automatic feature extraction without much human intervention, it is largely conceived as a ``black box," and the design of a neural network is typically a trial-and-error process. The transparency issue of the DNN has received increasing attention lately, where both machine learning and visualization communities have attempted to understand the training process and interpret the machine cognition in a visually conceivable way. Craven and Shavlik \cite{Craven1992vis} surveyed several visualization techniques, including Hinton diagrams, bond diagrams, hyperplane diagrams, response-function plots, and trajectory diagrams, which help provide visual evidence for understanding the learning and decision-making processes of neural networks. Mohamed {\it et al.} \cite{Mohamed2012dimension} introduced a dimension-reduction technique which shows that pre-training, fine-tuning, and nonlinear hidden layers are three major components that contribute to the superb acoustic recognition performance of deep belief networks (DBNs). Rauber {\it et al.} \cite{Rauber2017dimension} analyzed the DNNs by visualizing the low-dimensional projections of hidden layer activations, and provided insights into how the learned representations of samples have evolved during training. Bach {\it et al.} \cite{Bach2015heatmap} introduced the layer-wise relevance propagation (LRP) algorithm for identifying important pixels linked to a particular DNN prediction in a given input image. The significant features extracted by the training process were visualized in the pixel/input space. Samek {\it et al.} \cite{Samek2017heatmap} showed that LRP, as compared to the sensitivity analysis and the deconvolution method, can more accurately identify important pixels based on which the DNN prediction was made.

Applying the visual analytics techniques to understand the working mechanism of DNNs in wireless indoor localization has not been studied. The significance of this endeavor is twofold. First, providing visual evidence and interpretation of DNN workings in wireless indoor localization is by itself valuable, as the wireless signal patterns generally lack visual evidence and human intuition to interpret important features. Making DNN classifier interpretable also opens up new ways for model improvement and allows for analysis and comparison of various machine learning models for wireless indoor localization. Second, since the features in wireless localization applications are physical signatures of multiple-input multiple-output (MIMO) wireless links, an investigation and discussion from specifically wireless perspectives provides new insights on DNNs which cannot be directly inferred from other domain applications. Specifically, in this paper, we address the following issues: 1) How well has the DNN learned and been trained in device-free Wi-Fi indoor localization? 2) What critical features (wireless signatures) has the DNN learned to distinguish different classes in device-free Wi-Fi indoor localization? Toward these ends, we adopt several visualization techniques to provide quantitative and plausible explanations of DNN workings in this specific application. Our main contributions are summarized as follows:
\begin{itemize}
\item We model the device-free wireless indoor localization as a classification problem using the DNN, which yields a better average precision and recall rate than the $k$-NN and SVM techniques.
\item To the best of our knowledge, this paper presents the first study that provides visual analysis to understand the mechanism of DNN-based device-free wireless indoor localization. We employ two visual analytics techniques, namely, dimensionality reduction visualization and visual analytics and information visualization, to analyze the resultant clustering effect among the data after DNN and visualize the critical channels of a given input relative to a particular output decision.
\item We exploit the relevance score of a channel in an input CSI relative to a specific output class by regarding it as a correlation coefficient between two channels: one associated with the location class of the input CSI and the other associated with the specific output class that is learned by the DNN. Particularly, we develop the progressive channel nullification and channel modification procedures to evaluate the impacts of the channels with higher absolute values of relevance scores on the output prediction of each class. With extensive experiments using these two channel manipulation procedures, we justify that relevance scores are truly reliable quantitative indicators of critical features that distinguish classes, which allows for a better understanding of the critical features in wireless signals learned by the DNN which cannot be well interpreted by human perceptions.
\end{itemize}

The rest of the paper is organized as follows. Section~\ref{sec:experiments} provides preliminaries on the system and experiments of DNN-based device-free Wi-Fi indoor localization. Section~\ref{sec:method} presents the visual analytics techniques. Sections~\ref{sec:results_exp1} and \ref{sec:results_exp2} present the results and discussion. Finally, Section~\ref{sec:conclusion} concludes the paper.

\section{Experiments and Datasets} \label{sec:experiments}

Two real-world experiments for device-free Wi-Fi indoor localization were conducted, in a conference room and in a lounge, respectively, at the Research Center for Information Technology Innovation, Academia Sinica. The two environments exemplify different but common, realistic indoor settings (open space in the conference room and space with furniture/obstructions in the lounge) for our analysis.

\subsection{Experimental Settings}

\begin{figure}[t]
\begin{center}
\subfigure[]{
    \label{fig:exp1_floorplan}
    \includegraphics[width=0.48\columnwidth,height=2.2in]{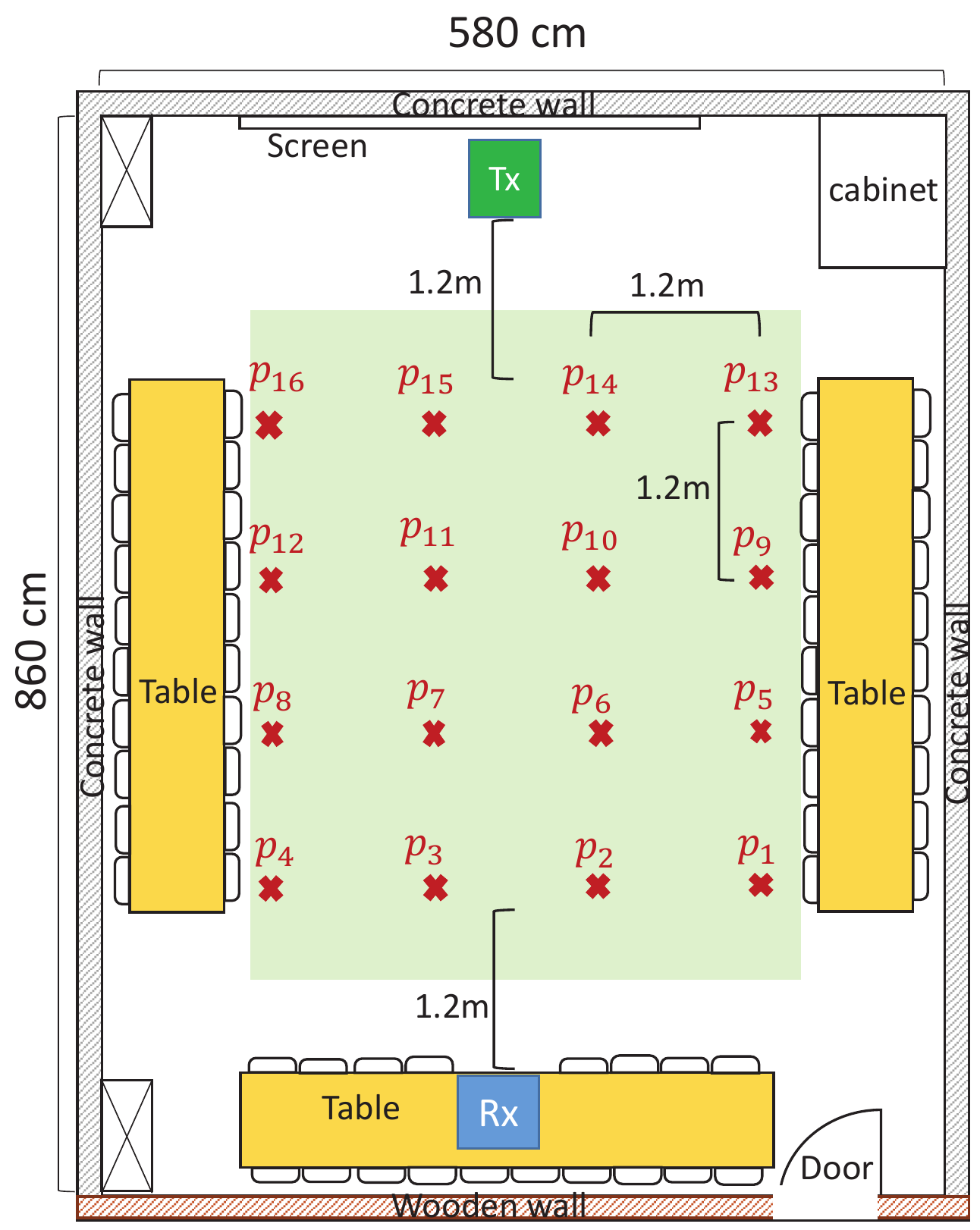}}
\hspace*{-0.06in}
\subfigure[]{
    \label{fig:exp1_photograph}
    \includegraphics[width=0.48\columnwidth,height=2.2in]{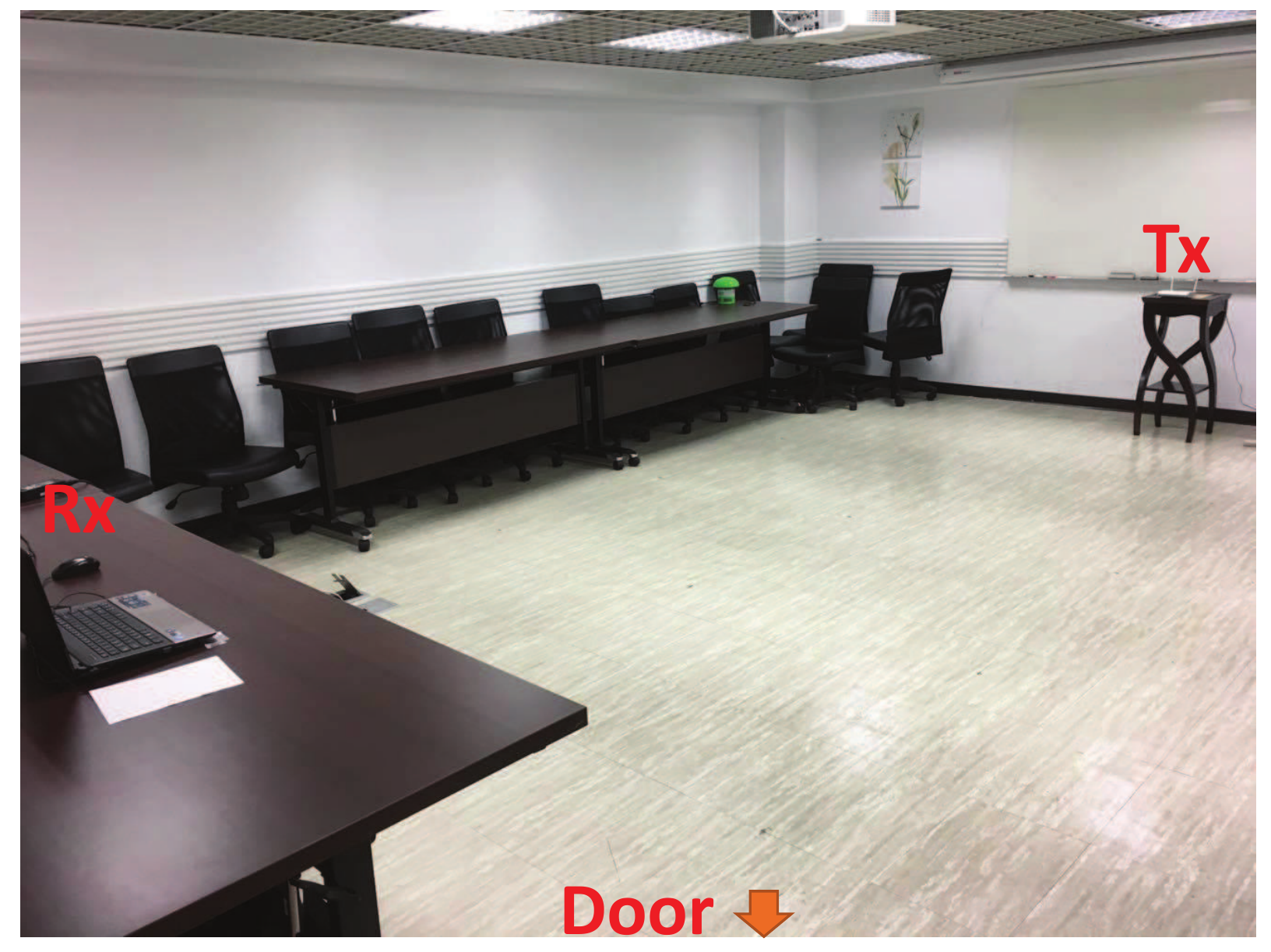}}
\caption{Experiment 1 (conference room). A single Wi-Fi AP (Tx) and a single fixed-location receiver (Rx) are deployed for device-free indoor localization. (a) Floor plan. (b) Photograph.}
\label{fig:exp1}
\end{center}
\end{figure}

\begin{figure}[t]
\begin{center}
\subfigure[]{
    \label{fig:exp2_floorplan}
    \includegraphics[width=0.48\columnwidth,height=2.2in]{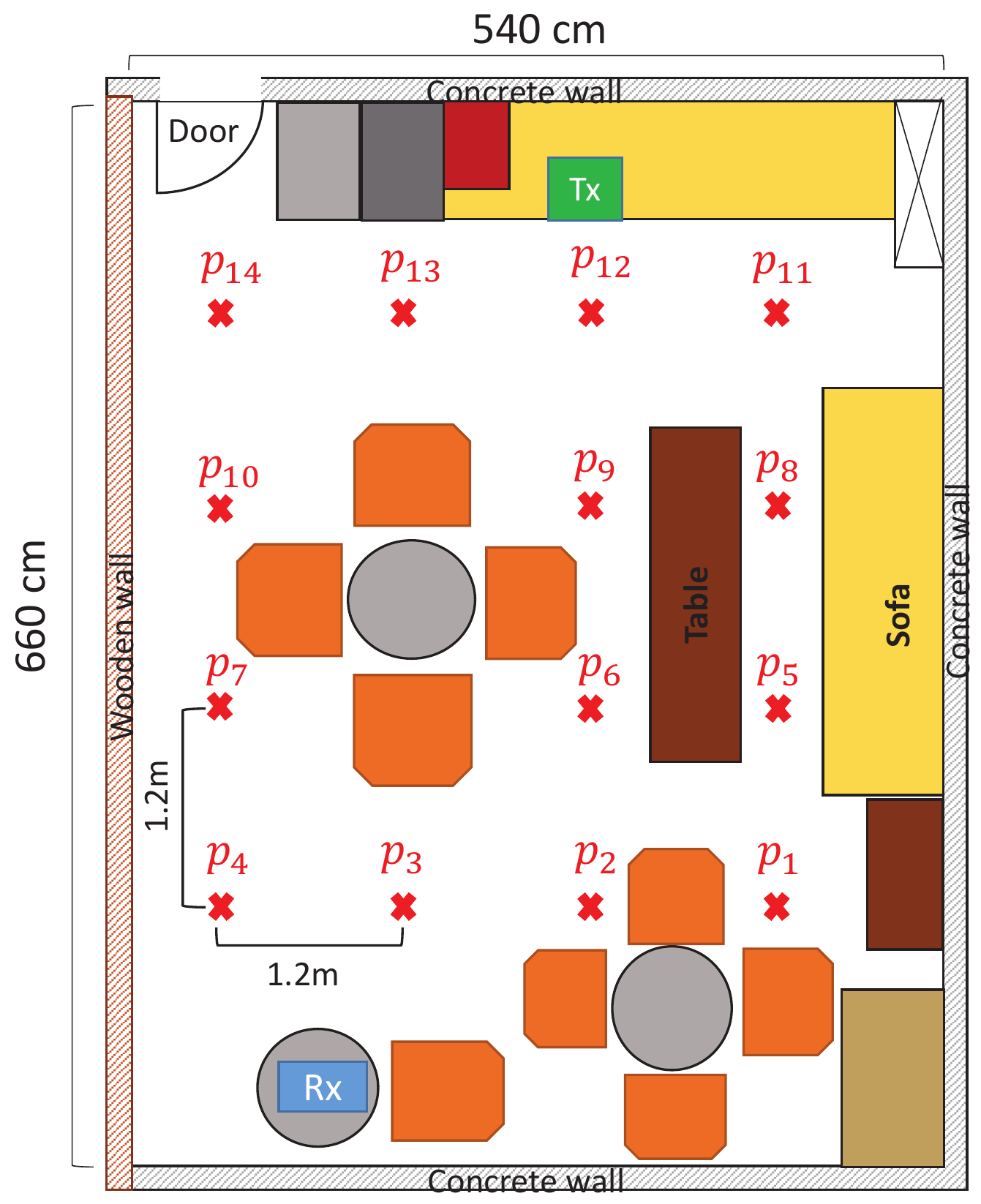}}
\hspace*{-0.06in}
\subfigure[]{
    \label{fig:exp2_photograph}
    \includegraphics[width=0.48\columnwidth,height=2.2in]{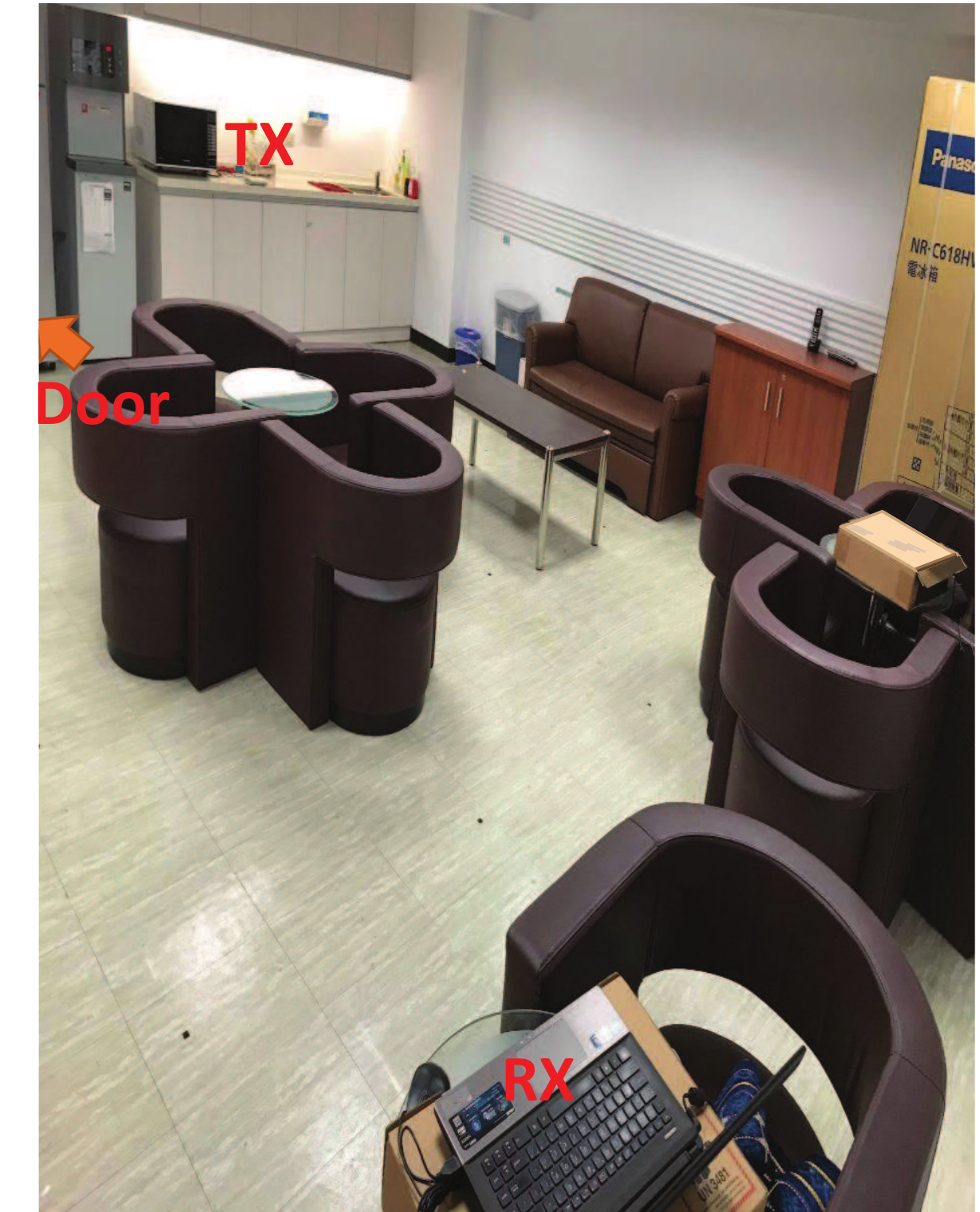}}
\caption{Experiment 2 (lounge). A single Wi-Fi AP (Tx) and a single fixed-location receiver (Rx) are deployed for device-free indoor localization. (a) Floor plan. (b) Photograph.}
\label{fig:exp2}
\end{center}
\end{figure}

The experimental settings in the conference room are depicted in Fig.~\ref{fig:exp1}. The conference room has dimensions $5.8\times 8.6$ m$^2$. There are $M=16$ target locations ($p_1,p_2,\ldots,p_{16}$) which are $1.2$ m apart between any two nearest locations. A single Wi-Fi access point (AP) of the model Zyxel NBG-419N as the transmitter (Tx) and a single ASUS laptop equipped with Ubuntu 10.04 LTS and Intel Wi-Fi Wireless Link 5300 802.11n MIMO-OFDM radios \cite{Daniel2011tool5300} as the receiver (Rx) are both placed at fixed locations with heights $88$ cm and $82$ cm, respectively. No tracking device is attached to the target to be localized (i.e., device-free). The receiver pings the packets sent from the transmitter and collects channel state information (CSI) during packet transmissions. The CSI data are used to train a DNN-based classification model for localization.

The experimental settings in the lounge are depicted in Fig.~\ref{fig:exp2}. The lounge has dimensions $5.4\times 6.6$ m$^2$. There are $M=14$ target locations ($p_1,p_2,\ldots,p_{14}$) which are $1.2$ m apart between two nearest locations. Unlike the conference room experiment, the target locations here are not uniformly arranged due to obstructions. The Wi-Fi transmitter and receiver are placed at fixed locations with heights $76$ cm and $69$ cm, respectively. All other settings are the same as in the conference room experiment.

\subsection{Datasets}

In the conference room experiment, CSI data for both training and testing phases were collected. In the offline training phase, the same person stood at each of the $16$ locations and then the fixed-location receiver pings the packets from the transmitter. The same procedure was repeated eight times, in two days and at four different times each day (morning, noon, afternoon, evening). Overall, $200$ CSI samples were collected for each location ($3200$ samples for all $16$ locations). Each CSI sample is a $K$-dimensional vector, where $K=120$ is specified by the number of OFDM subcarriers ($30$) multiplied by two transmit and two receive antennas. The elements of the $K$-dimensional CSI sample, termed {\it CSI channels} or simply {\it channels} in this paper, are indexed in the order of $30$ subcarriers for transmit antenna 1 and receive antenna 1 pairing (denoted by Tx1--Rx1), $30$ subcarriers for Tx1--Rx2, $30$ subcarriers for Tx2--Rx1, and $30$ subcarriers for Tx2--Rx2. In the online testing phase, a similar task was performed, which is independent of the training phase. Each of the $16$ locations was tested twice, at different times of the same day to incorporate environmental variations. In each test, $100$ CSI samples were collected for each location ($1600$ for all $16$ locations).

In the lounge experiment, in the offline training phase, the same person stood at each of the $14$ locations with four different orientations at each location (facing forward, backward, right, and left with respect to the transmitter), and the fixed-location receiver pings the packets from the transmitter. The same procedure was repeated $12$ times, in two days with six times of two hours apart each day. Overall, $384$ CSI samples were collected for each location ($5376$ samples for all $14$ locations). In the online testing phase, the same task was performed, independently of the training phase. Each of the $14$ locations was tested twice, at different times of the same day. In each test, $92$ CSI samples were collected for each location ($1288$ for all $14$ locations).

\subsection{Deep Neural Networks}

\begin{figure}[t]
\begin{center}
\includegraphics[width=0.85\columnwidth]{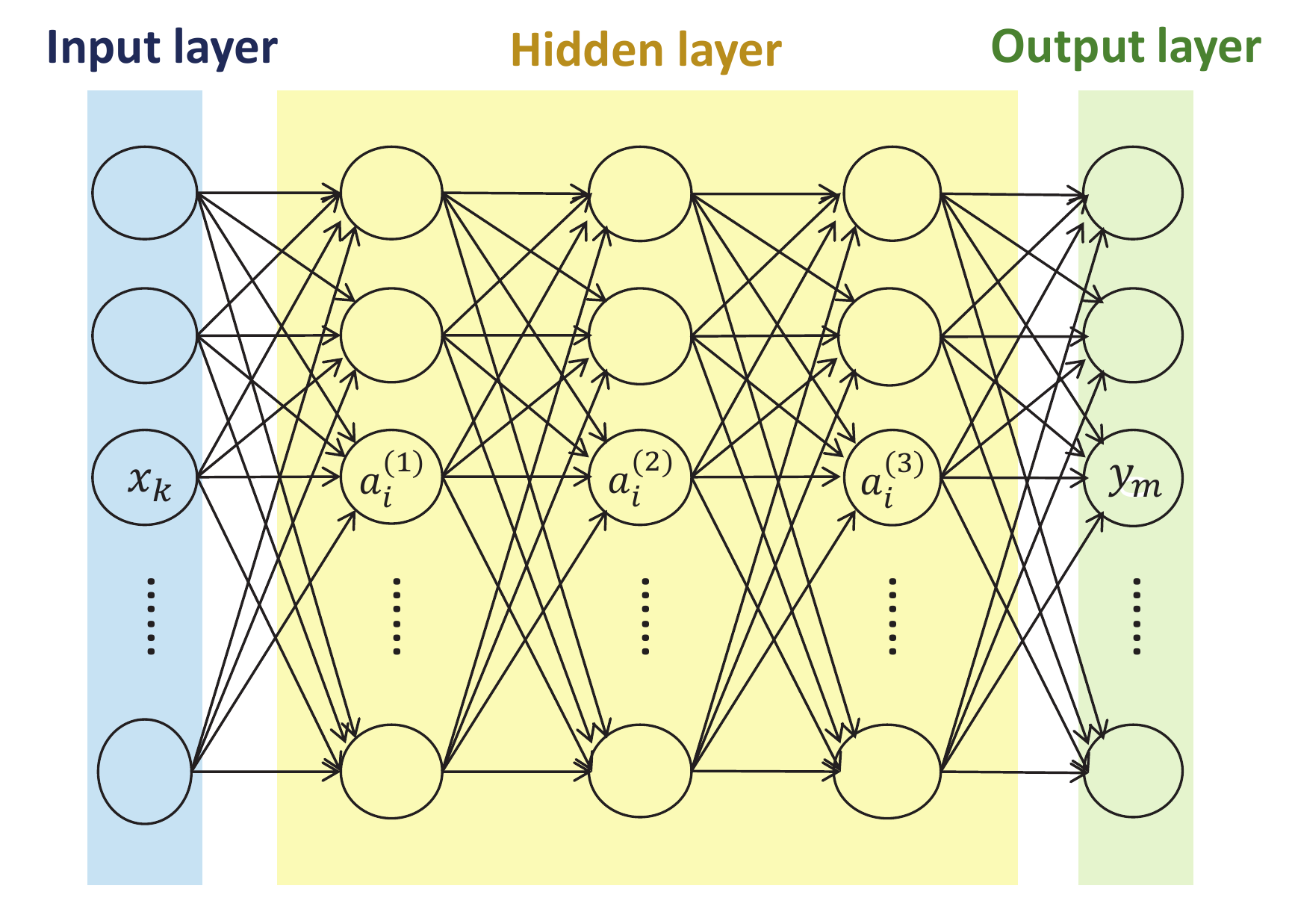}
\caption{The architecture of the DNN.}
\label{fig:DNN}
\end{center}
\end{figure}

A fully connected feedforward network consisting of an input layer, $L=3$ hidden layers (from first to third, $300$, $280$, and $260$ neurons, respectively)\footnote{It is known that by increasing the number of hidden neurons we can improve the approximation of the desired function, yet with a higher computational cost. The selection of the number of hidden layers and the number of neurons in each hidden layer here is based on both conventional wisdom and our experiments to strike a reasonable tradeoff between complexity and performance.}, and an output layer is adopted, as illustrated in Fig.~\ref{fig:DNN}. The input to the input layer is the $K$-dimensional raw CSI data, denoted by ${\mathbf x}=[x_1, x_2, \ldots, x_{K}]^T$, where $x_k$ denotes the input data of CSI channel $k$. The input to neuron $i$ in the first hidden layer is $z_i^{(1)}=\sum_{j} w_{i,j}^{(1)}x_{j} + b_{i}^{(1)}$, where $w_{i,j}^{(1)}$ and $b_{i}^{(1)}$ are the weight and bias with self-explanatory indices. The output of neuron $i$ in the first hidden layer is given by $a_{i}^{(1)}={\rm ReLU}\big( z_i^{(1)}\big)$, where ${\rm ReLU}(\cdot)$ is the rectified linear unit (ReLU) activation function defined by ${\rm ReLU}(x)=\max(0,x)$. The input of neuron $i$ in the $l$th ($l=2,\ldots,L$) hidden layer is similarly given by $z_{i}^{(l)}=\sum_{k=1}w_{i,k}^{(l)}a_{k}^{(l-1)}+b_{i}^{(l)}$, with the corresponding output $a_{i}^{(l)}={\rm ReLU}\big( z_{i}^{(l)} \big)$. The softmax output of neuron $i$ in the output layer is described by $y_i=\exp\big( z_i^{(L+1)} \big)/\sum_{m=1}^M \exp\big( z_m^{(L+1)}\big)$. The prediction is location $p_{m'}$, where $m'=\mathop{\arg\max}\limits_{m}~y_m\triangleq f({\bf x})$, with $f(\cdot)$ denoting the decision function of the DNN. The network is trained in a supervised manner with the cross-entropy objective function, with $30$ iterations of pre-training and $1500$ iterations of backpropagation.

\subsection{Performance}

\begin{table*}[t]
\begin{center}
\caption{Experiment 1 (Conference Room): Precision and Recall (Unit: $\%$, after Rounding) for Locations $p_{1},\ldots,p_{16}$ for $k$-NN, SVM, and DNN Based Device-Free Indoor Localization}
\label{tab:exp1_Precision and Recall} \vspace*{1mm}
\begin{tabular}{|l|l|l|l|l|l|l|l|l|l|l|l|l|l|l|l|l|l|l|} \hline
Measure & Scheme &$p_{1}$ & $p_{2}$ & $p_{3}$ & $p_{4}$ & $p_{5}$ & $p_{6}$ & $p_{7}$ & $p_{8}$ & $p_{9}$ & $p_{10}$ & $p_{11}$ & $p_{12}$ & $p_{13}$ & $p_{14}$ & $p_{15}$ & $p_{16}$ & Avg. \\\hline\hline
\multirow{3}{*}{Precision} & $k$-NN & $85$ & $100$ & $59$ & $100$ & $85$ & $99$ & $100$ & $95$ & $31$ & $70$ & $98$ & $73$ & $40$ & $98$ & $100$ & $83$ & $82.25$ \\\cline{2-19}
& SVM & $100$ & $100$ & $47$ & $96$ & $93$ & $91$ & $96$ & $97$ & $61$ & $79$ & $95$ & $93$ & $58$ & $100$ & $100$ & $70$ & $86$ \\\cline{2-19}
& DNN & $100$ & $100$ & $33$ & $94$ & $100$ & $96$ & $75$ & $85$ & $90$ & $76$ & $100$ & $94$ & $51$ & $100$ & $95$ & $88$ & $86.06$ \\\hline\hline
\multirow{3}{*}{Recall} & $k$-NN & $34$ & $45$ & $96$ & $53$ & $41$ & $100$ & $100$ & $72$ & $99$ & $96$ & $85$ & $38$ & $55$ & $60$ & $60$ & $100$ & $70.88$ \\\cline{2-19}
& SVM & $19$ & $100$ & $99$ & $78$ & $26$ & $100$ & $99$ & $55$ & $95$ & $98$ & $99$ & $100$ & $84$ & $53$ & $60$ & $100$ & $79.06$ \\\cline{2-19}
& DNN & $8$ & $100$ & $62$ & $96$ & $11$ & $100$ & $96$ & $65$ & $100$ & $100$ & $100$ & $100$ & $89$ & $55$ & $88$ & $100$ & $79.37$ \\\hline
\end{tabular}
\end{center}
\end{table*}

\begin{table*}[t]
\begin{center}
\caption{Experiment 2 (Lounge): Precision and Recall (Unit: $\%$, after Rounding) for Locations $p_{1},\ldots,p_{14}$ for $k$-NN, SVM, and DNN Based Device-Free Indoor Localization}
\label{tab:exp2_Precision and Recall} \vspace*{1mm}
\begin{tabular}{|l|l|l|l|l|l|l|l|l|l|l|l|l|l|l|l|l|} \hline
Measure & Scheme &$p_{1}$ & $p_{2}$ & $p_{3}$ & $p_{4}$ & $p_{5}$ & $p_{6}$ & $p_{7}$ & $p_{8}$ & $p_{9}$ & $p_{10}$ & $p_{11}$ & $p_{12}$ & $p_{13}$ & $p_{14}$  & Avg. \\\hline\hline
\multirow{3}{*}{Precision} & $k$-NN & $36$ & $97$ & $17$ & $62$ & $71$ & $43$ & $52$ & $74$ & $82$ & $51$ & $96$ & $100$ & $45$ & $46$ &  $62.29$ \\\cline{2-17}
& SVM & $98$ & $100$ & $24$ & $86$ & $68$ & $48$ & $74$ & $96$ & $98$ & $93$ & $66$ & $41$ & $52$ & $76$ & $72.86$ \\\cline{2-17}
& DNN & $94$ & $98$ & $34$ & $89$ & $61$ & $65$ & $92$ & $35$ & $66$ & $84$ & $88$ & $99$ & $53$ & $68$ & $73.29$ \\\hline\hline
\multirow{3}{*}{Recall} & $k$-NN & $16$ & $16$ & $65$ & $26$ & $65$ & $60$ & $36$ & $57$ & $58$ & $36$ & $14$ & $86$ & $67$ & $63$ & $47.5$ \\\cline{2-17}
& SVM & $28$ & $33$ & $65$ & $33$ & $65$ & $65$ & $63$ & $65$ & $51$ & $65$ & $51$ & $98$ & $64$ & $65$ & $57.93$ \\\cline{2-17}
& DNN & $43$ & $42$ & $65$ & $70$ & $65$ & $65$ & $61$ & $78$ & $63$ & $65$ & $29$ & $96$ & $80$ & $70$ & $63.71$ \\\hline
\end{tabular}
\end{center}
\end{table*}

Table~\ref{tab:exp1_Precision and Recall} and Table~\ref{tab:exp2_Precision and Recall} summarize the location-specific statistical measures, i.e., {\it precision} and {\it recall}, for the conference room experiment and lounge experiment, respectively, for DNN-based device-free indoor localization system in comparison with $k$-NN ($k=5$, the best-performing configuration\footnote{More precisely, $k=5$ achieves the highest macro-average recall and second-highest macro-average precision in the conference room experiment, and the highest macro-average recall and precision in the lounge experiment, among commonly used $k=1, 3, 5, 7$ values. Besides, $k=5$ achieves the highest F1-score, which is the harmonic mean of the macro-average precision and recall, in both experiments.}) and SVM. $k$-NN classification is based on the raw CSI data and is adopted to benchmark the analysis of DNN in Sections~\ref{sec:results_exp1} and \ref{sec:results_exp2}. SVM with a Gaussian radial basis function (RBF) kernel and the one-against-all technique \cite{Zhou2017ml,Wu2016} for solving the multi-classification problem is adopted. The precision and recall measure the percentages of correct classification for each predicted and true class in the testing, respectively. 

\section{Visual Analytics Techniques} \label{sec:method}

Two visual analytics techniques are employed to facilitate the analysis, visualization, and evaluation of the neural network in the application of device-free Wi-Fi indoor localization.

\subsection{Dimensionality Reduction Visualization} 

A dimensionality reduction technique called t-distributed stochastic neighbor embedding (t-SNE) \cite{Maaten2008tsne} is employed to project the high-dimensional data to the low-dimensional space for ease of visualization. In the projection, t-SNE aims to preserve the pairwise relationship of the high-dimensional data in the low-dimensional space. Specifically, t-SNE first converts the pairwise distances between datapoints in the high-dimensional space to joint probabilities, i.e., the joint probabilities of high-dimensional datapoints ${\mathbf x}_i$ and ${\mathbf x}_j$ are defined as $p_{ij}=\exp(-\left\|{\mathbf x}_i-{\mathbf x}_j\right\|^2)/\sum_{k\neq l}\exp(-\left\|{\mathbf x}_k-{\mathbf x}_l\right\|^2)$ for $i\neq j$, and $p_{ii}=0,\forall i$. Then, t-SNE finds the low-dimensional data representation such that the low-dimensional counterparts have similar joint probabilities as the high-dimensional data. The joint probabilities $q_{ij}$ for the low-dimensional counterparts $\boldsymbol{\chi}_i$ and $\boldsymbol{\chi}_j$ are defined as $q_{ij}=(1+\left\|\boldsymbol{\chi}_i-\boldsymbol{\chi}_j\right\|^2)^{-1}/\sum_{k\neq l} (1+\left\|\boldsymbol{\chi}_k-\boldsymbol{\chi}_l\right\|^2)^{-1}$ for $i\neq j$, and $q_{ii}=0,\forall i$, which is based on the heavy-tailed Student t-distribution instead of Gaussian as in the high-dimensional space to avoid the ``crowding problem" \cite{Maaten2008tsne} in the low-dimensional representation. t-SNE finds the low-dimensional data representation with such $q_{ij}$'s that the Kullback-Leibler divergence between $p_{ij}$'s and $q_{ij}$'s, i.e., $\sum_i \sum_j p_{ij}\log (p_{ij}/q_{ij})$, is minimized.

In our application, the high-dimensional input or hidden-layer activations are projected to the 2D space via t-SNE and presented as scatterplots. In the scatterplot, the points are colored or labeled according to the corresponding classes (i.e., locations $p_1,\ldots,p_{16}$). We adopt the {\it silhouette score} \cite{Rousseeuw1987SC} to visually assess the clustering of points into different classes in the 2D representation. The silhouette score $s(i)$ of the $i$th point is defined as $s(i)=(d_{\rm nearest}(i)-d_{\rm intra}(i)) / \max\{d_{\rm intra}(i), d_{\rm nearest}(i)\}$, where $d_{\rm intra}(i)$ is the mean distance between the $i$th point and all other points within the same cluster (i.e., the mean intra-cluster distance), and $d_{\rm nearest}(i)$ is the minimum mean distance between the $i$th point and all points in any other cluster (i.e., the mean nearest-cluster distance). The silhouette score lies in the interval $[-1, 1]$. The score is close to $1$ if $d_{\rm nearest}(i) \gg d_{\rm intra}(i)$, i.e., the data are well-clustered, close to $-1$ if $d_{\rm nearest}(i) \ll d_{\rm intra}(i)$, i.e., the data are not well-clustered in the sense that the data would have been better clustered if clustered to the neighboring cluster, and near zero if $d_{\rm nearest}(i) \approx d_{\rm intra}(i)$, i.e., the data are in between two clusters. We calculate the average silhouette score over all points for each scatterplot.

\subsection{Visual Analytics and Information Visualization} 

Layer-wise relevance propagation (LRP) \cite{Bach2015heatmap} is developed to quantify the contribution of the data in each input neuron  to a specific output prediction. This is achieved by decomposing a specific pre-softmax value, i.e., $z_{m}^{(L+1)}$, in the output layer into {\em relevance scores} of the neurons in the previous hidden layer which are then propagated backward toward the input layer. The back-propagation rule in LRP is constrained by the conservation principle \cite{Bach2015heatmap}, i.e.,
\begin{align} 
\sum_i h_i(p_n\to p_m) &= \sum_i R_i^{(l)}(p_n\to p_m) = z_{m}^{(L+1)}, \nonumber\\
& m,n=1,\ldots,M; l=1,\ldots,L \label{eq:R_conservation}
\end{align}
where $h_i(p_n\to p_m)$ and $R_i^{(l)}(p_n\to p_m)$ are the relevance scores of CSI channel $i$ in the input layer and of neuron $i$ in the $l$th hidden layer, respectively, corresponding to the input CSI data of location $p_n$ and the output DNN decision of location $p_m$. The conservation principle forces the sum of the relevance scores to be preserved in all layers and equal to $z_{m}^{(L+1)}$, the output score before softmax. The conservation principle can be applied repeatedly backward in all layers to obtain the relevance scores. The relevance score of neuron $i$ in the $L$th hidden layer is given by
\begin{align}
R_i^{(L)}(p_n\to p_m) = \frac{a_{i}^{(L)}w_{m,i}^{(L+1)}}{\sum_{j}a_{j}^{(L)}w_{m,j}^{(L+1)}+b_{m}^{(L+1)}}z_m^{(L+1)}
\end{align}
for some $m, n$, which approximates the conservation principle. The relevance scores of neuron $i$ in the $l$th ($l=1,\ldots,L-1$) hidden layer and of CSI channel $i$ in the input layer are given respectively by
\begin{align}
R_i^{(l-1)}(p_n\to p_m) &= \sum_{k}\frac{a_{i}^{(l-1)}w_{k,i}^{(l)}}{\sum_{j}a_{j}^{(l-1)}w_{k,j}^{(l)}+b_{k}^{(l)}}R_k^{(l)}(p_n\to p_m), \nonumber\\ 
&\qquad\qquad\qquad\qquad\qquad l=2,\ldots,L, \label{eq:R_layer} \\
h_i(p_n\to p_m) &= \sum_{k}\frac{x_{i}w_{k,i}^{(1)}}{\sum_{j}x_{j}w_{k,j}^{(1)}+b_{k}^{(1)}}R_k^{(1)}(p_n\to p_m)\label{eq:R_input}
\end{align}
for some $m, n$. The relevance score $h_i(p_n\to p_m)$ is normalized to $[-1,1]$ by $h'_i(p_n\to p_m)=\frac{h_i(p_n\to p_m)}{\max_{j=1,\ldots,K}(\left|  h_j(p_n\to p_m) \right|)}$. A positive valued $h'_i(p_n\to p_m)$ suggests that CSI channel $i$ provides positive evidence in support of the DNN prediction of location $p_m$ given the input CSI data of location $p_n$, a negative valued $h'_i(p_n\to p_m)$ suggests that CSI channel $i$ provides negative evidence against such a prediction, and a near-zero $h'_i(p_n\to p_m)$ suggests that CSI channel $i$ provides little evidence for or against such a prediction.

To examine how each input CSI channel has contributed to a specific output DNN decision through the relevance scores, we propose two channel manipulation processes, inspired by the approach in \cite{Bach2015heatmap} where the process is associated with flipping the pixel value of a binary image. We first define, for some $m,n$, the following channel ordering sequences, where each $(r_1,\ldots, r_{K})$ represents a permutation of the natural channel ordering $(1,\ldots,K)$ according to a certain ordering of the normalized relevance scores:
\begin{enumerate}
\item $\mathcal{O}_1(p_n{\to} p_m) = (r_1,\ldots, r_{K})$ such that the corresponding sequence of relevance scores is in descending order, i.e., $h'_{r_1}(p_n\to p_m)\geq h'_{r_2}(p_n\to p_m)\geq\cdots\geq h'_{r_K}(p_n\to p_m)$.
\item $\mathcal{O}_2(p_n{\to} p_m) = (r_1,\ldots, r_{K})$ such that the corresponding sequence of relevance scores is in ascending order, i.e., $h'_{r_1}(p_n\to p_m)\leq h'_{r_2}(p_n\to p_m)\leq\cdots\leq h'_{r_K}(p_n\to p_m)$.
\item $\mathcal{O}_3(p_n{\to} p_m) = (r_1,\ldots, r_{K})$ such that the corresponding sequence of relevance scores satisfies $|h'_{r_1}(p_n\to p_m)|\geq |h'_{r_2}(p_n\to p_m)|\geq\cdots\geq |h'_{r_K}(p_n\to p_m)|$.
\item $\mathcal{O}_4(p_n{\to} p_m) = (r_1,\ldots, r_{K})$ such that the corresponding sequence of relevance scores satisfies $|h'_{r_1}(p_n\to p_m)|\leq |h'_{r_2}(p_n\to p_m)|\leq\cdots\leq |h'_{r_K}(p_n\to p_m)|$.
\end{enumerate}

Then, the proposed channel manipulation processes to facilitate information visualization are described below.
\subsubsection{Channel Nullification} The process is to progressively nullify or remove the channel in the input CSI data of location $p_n$, denoted by ${\mathbf x}_{p_n}=[x_{p_n,1},x_{p_n,2},\ldots,x_{p_n,K}]^T$, one channel at a time according to a specific channel ordering sequence, and measure the impact of the nullified channel on the classification result. The $t$th ($t=1,\ldots,K$) step in the removal process returns ${\mathbf x}_{p_n}^{(t)}=g_{\rm null}({\mathbf x}_{p_n}^{(t-1)},r_t)$, where ${\mathbf x}_{p_n}^{(0)}\triangleq {\mathbf x}_{p_n}$, and the function $g_{\rm null}$ sets the $r_t$-th element of ${\mathbf x}_{p_n}^{(t-1)}$ to zero. The influence of nullified CSI channels is reflected by the classification decision $f({\mathbf x}_{p_n}^{(t)})$ of the DNN.
\subsubsection{Channel Modification} The process is to progressively modify the channel in the input CSI data of location $p_n$ ``toward'' that of location $p_m$, one channel at a time according to a specific channel ordering sequence, given the input CSI data of location $p_n$ and the output DNN decision of location $p_m$. The $t$th ($t=1,\ldots,K$) step in the modification process returns ${\mathbf x}_{p_n}^{(t)}=g_{\rm mod}({\mathbf x}_{p_n}^{(t-1)},r_t)$, where ${\mathbf x}_{p_n}^{(0)}\triangleq {\mathbf x}_{p_n}$, and the function $g_{\rm mod}$ sets the $r_t$-th element of ${\mathbf x}_{p_n}^{(t-1)}$ to $\max\left(0, {\mathbb E}\left[x_{p_m,r_t}\right]{+}h'_{r_t}(p_n{\to} p_m) \frac{\sigma_{x_{p_m,r_t}}}{\sigma_{x_{p_n,r_t}}} \left(x_{p_n,r_t} {-} {\mathbb E}\left[x_{p_n,r_t}\right] \right)\right)$, which is based on the linear minimum mean square error (LMMSE) estimation of unknown $X$ given $Y$, i.e., $\widehat{X}={\mathbb E}[X]+\rho \frac{\sigma_X}{\sigma_Y}(Y - {\mathbb E}[Y])$, with correlation coefficient $\rho$ and standard deviations $\sigma_X,\sigma_Y$. The influence of modified CSI channels is reflected by the classification decision $f({\mathbf x}_{p_n}^{(t)})$ of the DNN.

We examine the percentage of correct classification for each true class $p_n$ in the testing (i.e., recall), and also the percentage of classification as location $p_m$ given the true class $p_n$, after channel nullification/modification. More specifically, we calculate, for $t=1,\ldots,K$, the number of $f({\mathbf x}_{p_n}^{(t)})=n$ and $f({\mathbf x}_{p_n}^{(t)})=m$, respectively, divided by the number of testing samples for location $p_n$.

The proposed technique can be extended to examine the contribution of each subcarrier to the DNN decision. We can calculate
\begin{align}
&s'_i(p_n\to p_m) \nonumber\\
&\quad=\frac{1}{4}\Big(h'_i(p_n\to p_m)+h'_{i+30}(p_n\to p_m) \nonumber\\
&\qquad +h'_{i+60}(p_n\to p_m)+h'_{i+90}(p_n\to p_m)\Big), \nonumber\\
&\qquad\qquad\qquad\qquad\qquad\qquad i=1,\ldots,30 \label{eq:subcarrier_relevance_score}
\end{align}
to represent the average normalized relevance scores of subcarrier $i$ corresponding to the input CSI data of location $p_n$ and output DNN decision of location $p_m$. Once we have the relevance scores of all the subcarriers corresponding to input $p_n$ and output $p_m$, we can define subcarrier ordering sequences similar to $\mathcal{O}_1$--$\mathcal{O}_4$ to rank the importance of subcarriers according to $s'_i(p_n\to p_m)$. By a slight abuse of notations, we use the same $\mathcal{O}_1$--$\mathcal{O}_4$ to denote the subcarrier ordering sequences and clearly make the distinction from channel ordering sequences in the context. Subcarrier nullification/modification procedures are also similarly defined. For example, nullification of subcarrier $i$ represents setting the $i$th, $(i+30)$th, $(i+60)$th, and $(i+90)$th elements of the CSI sample to zero.

\begin{figure*}[t]
\begin{center}
\includegraphics[width=2\columnwidth]{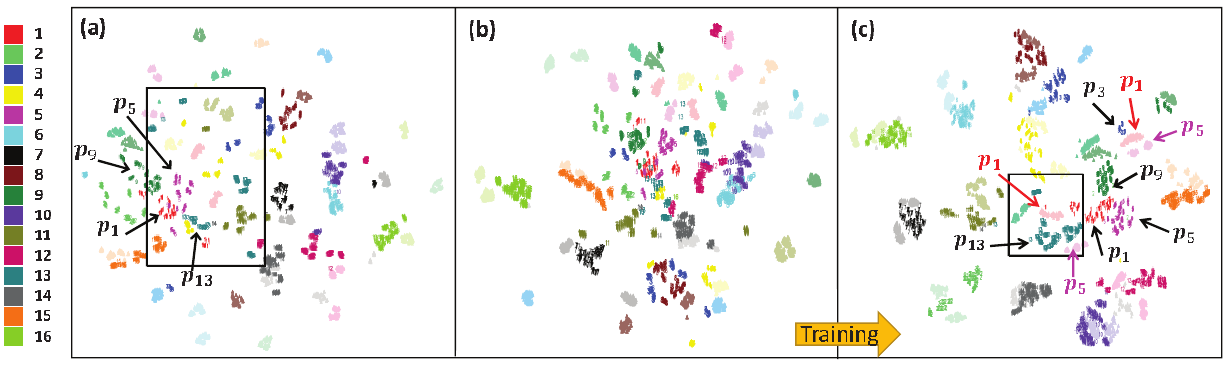}
\caption{The 2D visualization of (a) the raw CSI samples, (b) the last DNN hidden layer activations before training (with random initializations), and (c) the last DNN hidden layer activations after training. For each location, the training samples are shown with a darker shade to distinguish from the testing samples with a lighter shade of the same color. The silhouette scores (calculated for the training samples only) for (a)--(c) are $0.22$, $0.09$, and $0.66$, respectively. The rectangular boxes in (a) and (c) enclose all the training samples for location $p_{13}$. Black location markers refer to training samples and colored location markers (in colors corresponding to respective locations) refer to testing samples collected at the respective locations.}
\label{fig:tsne_training}
\end{center}
\end{figure*}

\begin{figure*}[t]
\begin{center}
\includegraphics[width=2\columnwidth]{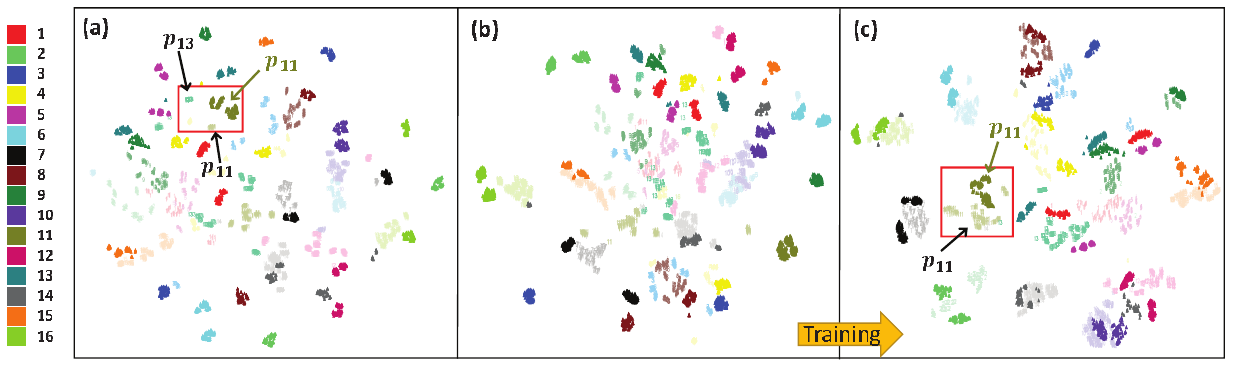}
\caption{Same plots as Fig.~\ref{fig:tsne_training}, but here, for each location, the testing samples are shown with a darker shade to distinguish from the training samples with a lighter shade of the same color. The silhouette scores (calculated for the testing samples only) for (a)--(c) are $-0.01$, $0.16$, and $0.41$, respectively. The rectangular boxes in (a) and (c) enclose the testing samples for location $p_{11}$ and the nearby training samples. Black location markers refer to training samples and colored location markers (in colors corresponding to respective locations) refer to testing samples collected at the respective locations.}
\label{fig:tsne_testing}
\end{center}
\end{figure*}

\section{Analysis and Visualization of Experimental Results: Experiment 1 (Conference Room)} \label{sec:results_exp1}

The results for experiment 1 (conference room) and experiment 2 (lounge) are analyzed and discussed in this section and next section, respectively.

\subsection{Training Effects} \label{sec:results_exp1_1}

The 2D visualization (via t-SNE) of 1) the raw CSI samples, 2) the last DNN hidden layer activations (i.e., $a_i^{(L)}$'s) before training (with random initializations, i.e., all the weights and biases are randomly generated from a Gaussian distribution with mean zero and standard deviation one), and 3) the last DNN hidden layer activations after training, are presented in Fig.~\ref{fig:tsne_training}(a)--(c), respectively. $200$ training samples and $100\times 2=200$ testing samples per location are plotted and color-coded. For each location, the training samples are shown with a darker shade to distinguish from the testing samples with a lighter shade of the same color. It is observed from Fig.~\ref{fig:tsne_training}(a) that the raw CSI samples are highly scattered, with a silhouette score of $0.22$, showing that the indoor localization problem at hand is not an easy classification problem. The untrained DNN shows similar visual separation among different classes (silhouette score: $0.09$) in Fig.~\ref{fig:tsne_training}(b). The trained DNN yields much improved visual separation (silhouette score: $0.66$), as shown in Fig.~\ref{fig:tsne_training}(c), by automatically extracting important CSI features to distinguish among different classes. It is expected that, as more distinctive patterns are learned through DNN training, different locations can be better classified with DNN in the testing. We use location $p_{13}$ as an example. The rectangular boxes in Fig.~\ref{fig:tsne_training}(a) and Fig.~\ref{fig:tsne_training}(c) enclose all the training samples for location $p_{13}$. It is seen that after DNN training these training samples become more concentrated. This leads to reduced misclassification of other locations as location $p_{13}$ in the testing. Specifically, for the DNN scheme, only locations $p_{1}$, $p_{5}$, and $p_{13}$ are classified as location $p_{13}$ in the testing, while for the $k$-NN scheme based on the raw CSI samples, locations $p_{1}$, $p_{4}$, $p_{5}$, $p_{9}$, $p_{12}$, $p_{13}$, and $p_{15}$ are classified as location $p_{13}$ in the testing. Quantitatively, DNN yields a higher precision than $k$-NN for location $p_{13}$ ($51\%$ vs. $40\%$), as shown in Table~\ref{tab:exp1_Precision and Recall}. Overall, DNN generally yields higher precision values, with a higher macro-average precision as compared to $k$-NN ($86.06\%$ vs. $82.25\%$). Thus, DNN training is effective in learning discriminative features from the original noisy wireless signals.

\subsection{Testing Performance} \label{sec:results_exp1_2}

Fig.~\ref{fig:tsne_testing} depicts the same figure as Fig.~\ref{fig:tsne_training}, but with the testing samples (instead of the training samples) highlighted in a darker shade of the same color. It is observed that Fig.~\ref{fig:tsne_testing}(c) presents better visual separation than Fig.~\ref{fig:tsne_testing}(a), with silhouette scores (calculated for the testing samples only) of $0.41$ vs. $-0.01$. Since testing was performed in two different time slots, Fig.~\ref{fig:tsne_testing}(c) suggests that the trained DNN model modifies the testing samples collected in different time slots to be more coherent, even though the raw testing data are more scattered in the data space due to environmental variations in two different time slots. The more concentrated distribution of the testing samples per location leads to the reduced misclassification of other locations as the true location in the testing, as shown by the higher macro-average recall for DNN than for $k$-NN ($79.37\%$ vs. $70.88\%$ in Table~\ref{tab:exp1_Precision and Recall}). 

Consider location $p_{11}$ as an example. In Fig.~\ref{fig:tsne_testing}(a), the testing samples for location $p_{11}$ are located near the training samples for locations $p_{11}$ and $p_{13}$, as enclosed by the rectangular box. After the DNN operation, the distribution of the testing data is altered, where the testing samples for location $p_{11}$ are now located near the training samples for location $p_{11}$ only, as enclosed by the rectangular box in Fig.~\ref{fig:tsne_testing}(c). The visual distribution matches the classification results where the recall for $k$-NN and DNN for location $p_{11}$ are $85\%$ and $100\%$, respectively.

\subsection{Line-of-Sight Effects} \label{sec:results_exp1_3}

It is observed from Table~\ref{tab:exp1_Precision and Recall} that, for DNN, $p_{1}$ and $p_{5}$ have very small values of recall but large values of precision. This suggests that location $p_{1}$ (or $p_{5}$) is largely misclassified as another location (small recall), and very few other locations are misclassified as location $p_1$ (or $p_{5}$) (large precision). A confusion matrix analysis reveals that the testing samples for location $p_{1}$ are predominantly misclassified as locations $p_3$ and $p_{13}$, and the testing samples for location $p_{5}$ are predominantly misclassified as locations $p_3$, $p_9$, and $p_{13}$. Fig.~\ref{fig:tsne_training}(c) and Fig.~\ref{fig:tsne_testing}(c) show that the testing samples for locations $p_{1}$ and $p_{5}$ are near the training samples for locations $p_{3}$, $p_{9}$, and $p_{13}$. Fig.~\ref{fig:tsne_training}(c) also shows that the training samples for locations $p_{1}$, $p_{5}$, $p_{9}$, and $p_{13}$ are close to each other, which suggests that locations $p_{1}$, $p_{5}$, $p_{9}$, and $p_{13}$ are not discernable after DNN learning. Fig.~\ref{fig:tsne_training}(a) shows that the raw CSI training samples for these four locations are also not well separated, which indicates that these original CSI patterns in the radio map are not discernable, making it difficult for the DNN to distinguish the location-specific features in the learning process. This may be in part because $p_{1}$, $p_{5}$, $p_{9}$, and $p_{13}$ are farther away from the line-of-sight (LoS) between the Wi-Fi transmitter and receiver. However, since the propagation of wireless signals is sensitive to room layouts and object placement, the CSI patterns and the resulting classification results are not the same for $p_{4}$, $p_{8}$, $p_{12}$, and $p_{16}$, which are also farther away from the LoS, in this experiment.

\subsection{Discriminative Features} \label{sec:results_exp1_4}

\begin{figure}[t]
\begin{center}
\includegraphics[width=1\columnwidth]{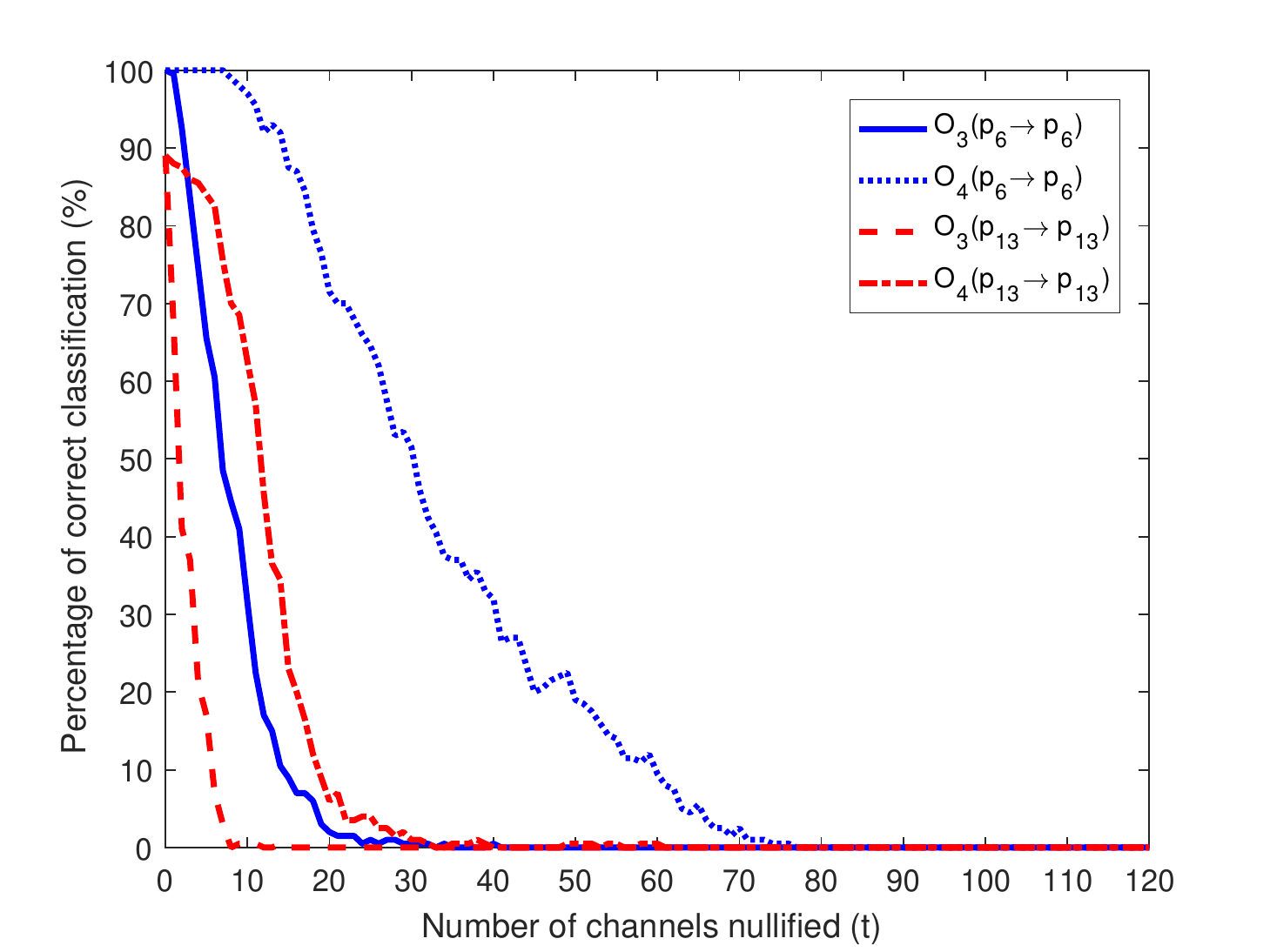}
\caption{The percentage of correct classification (true class: $p_6$ (or $p_{13}$)) after progressive channel nullification according to the channel ordering sequences $\mathcal{O}_3(p_6\to p_6)$ and $\mathcal{O}_4(p_6\to p_6)$ (or $\mathcal{O}_3(p_{13}\to p_{13})$ and $\mathcal{O}_4(p_{13}\to p_{13})$).}
\label{fig:LRP_p6_p13}
\end{center}
\end{figure}

\begin{figure}[t]
\begin{center}
\includegraphics[width=1\columnwidth]{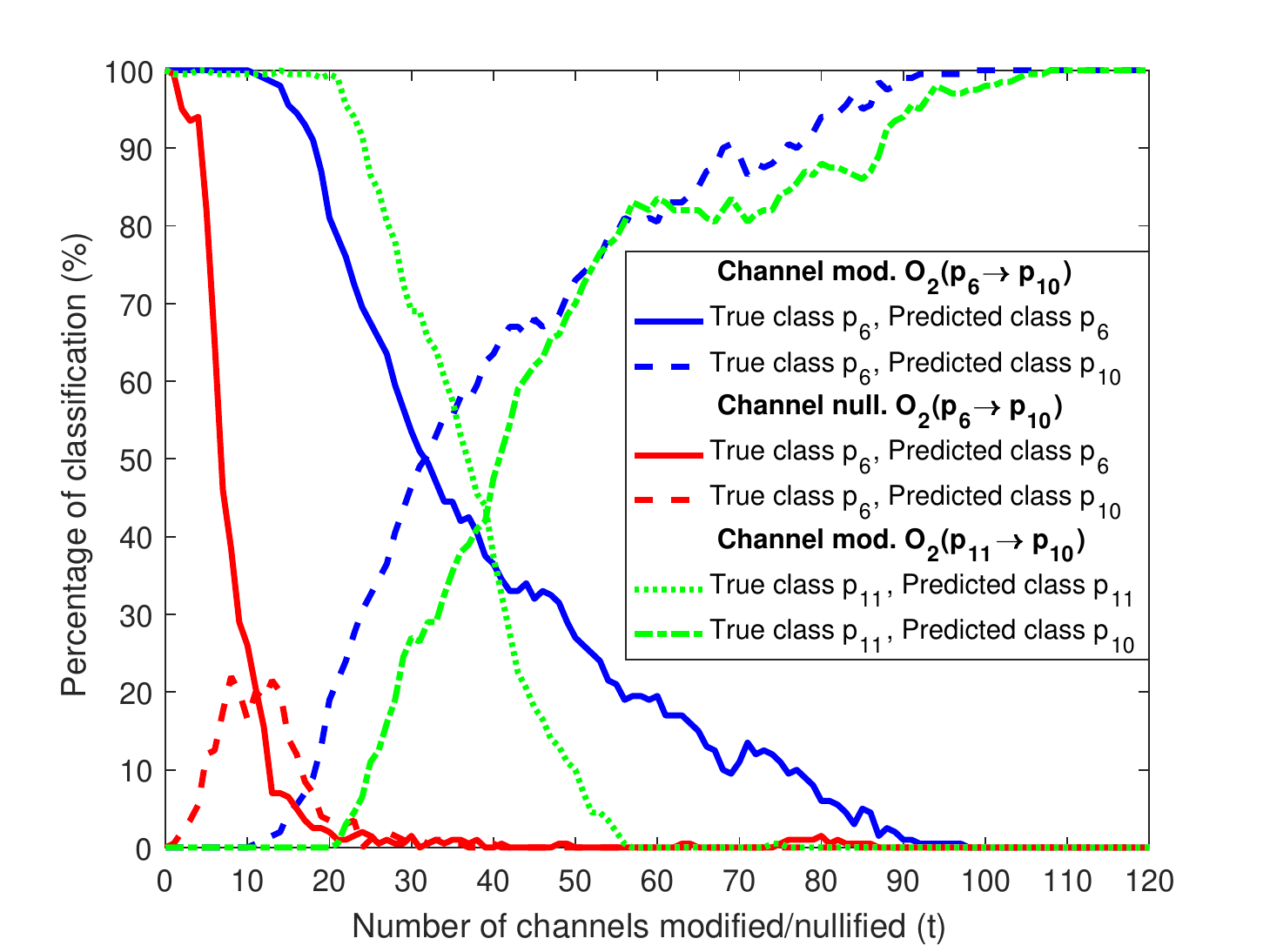}
\caption{The percentage of classification as $p_6$ and $p_{10}$ (true class: $p_6$) after progressive channel modification according to the channel ordering sequence $\mathcal{O}_2(p_6\to p_{10})$, in comparison with the same setting but after progressive channel nullification, and with a different pair of locations $p_{11}$ and $p_{10}$ (true class: $p_{11}$) after progressive channel modification according to the channel ordering sequence $\mathcal{O}_2(p_{11}\to p_{10})$.}
\label{fig:LRP_p6_p10_p11}
\end{center}
\end{figure}

Fig.~\ref{fig:LRP_p6_p13} shows the effect of progressive channel nullification on the DNN prediction using locations $p_6$ and $p_{13}$ as an example of locations nearer and farther from the LoS, respectively. For location $p_6$, channel nullification according to channel ordering sequences $\mathcal{O}_3(p_6\to p_6)$ and $\mathcal{O}_4(p_6\to p_6)$ represent, respectively, that the most relevant and irrelevant channels, as determined by LRP, leading to the correct DNN decision of location $p_6$ are progressively nullified. As can be seen, the percentage of correct classification generally decreases, more rapidly for $\mathcal{O}_3$ than for $\mathcal{O}_4$, as more channels are nullified. For $\mathcal{O}_3$, when as few as $20$ channels (out of $120$) are nullified, the percentage of correct classification drops from $100\%$ (the recall for location $p_6$) to nearly $0\%$, showing that the channels with large absolute values of relevance scores are indeed important and indispensable features for correct classification. Also, this shows that DNN can extract a small fraction of channels as dominant features. For $\mathcal{O}_4$, when the most irrelevant channels are nullified progressively, the percentage of correct classification still generally decreases, albeit more slowly. This is because, while these channels have little impact on the classification of the true class, they could have a pronounced impact on the classification of other classes. For example, when the channel corresponding to the leading element $r_1$ of $\mathcal{O}_4(p_6\to p_6)$ has a large CSI amplitude and a negative valued $h'_{r_1}(p_6\to p_{m})$ for some $m\neq 6$, nullifying this specific channel may erase the original negative evidence against the wrong class $p_m, m\neq 6$, resulting in an increased chance of a wrong DNN decision of location $p_m, m\neq 6$. For location $p_{13}$, the same general trend is observed, even though the percentage of correct classification drops even more rapidly from $89\%$ (the recall for location $p_{13}$) to $0\%$, since locations farther from the LoS could easily mutually misclassify as each other, as discussed previously.

Fig.~\ref{fig:LRP_p6_p10_p11} shows how the percentages of classification change as channels of the true class ($p_6$) are progressively modified toward another class ($p_{10}$) according to the channel ordering sequence $\mathcal{O}_2(p_6\to p_{10})$. Note that the leading elements of $\mathcal{O}_2(p_6\to p_{10})$ are channels that have the most negative relevance scores, i.e., provide the most negative evidence against the prediction of location $p_{10}$ given the true class being location $p_6$. By modifying these channels of location $p_6$ toward those of location $p_{10}$, we gradually erase the evidence against the prediction of location $p_{10}$ and enhance the evidence for the prediction of location $p_{10}$ (analogous to pixel flipping in binary images). Thus, we expect that the percentage of predicting location $p_6$ ($p_{10}$) will decrease (increase), provided that the relevance scores are reliable indicators of critical features that distinguish between classes. This is confirmed in Fig.~\ref{fig:LRP_p6_p10_p11}. In comparison, nullifying, instead of modifying, the same channels in the same order does not yield the same effect. Since channel nullification erases the evidence against, but does not enhance the evidence for, the prediction of location $p_{10}$, misclassification favors not only location $p_{10}$ but other locations as well. Fig.~\ref{fig:LRP_p6_p10_p11} also shows that channel nullification yields a significantly faster decreasing rate of correct classification than channel modification. This suggests that making radical changes to the channels such as nullification poses a more harmful effect on the correct classification than modifying the channels to another, possibly similar, pattern.

\begin{figure}[t]
\begin{center}
\includegraphics[width=\columnwidth]{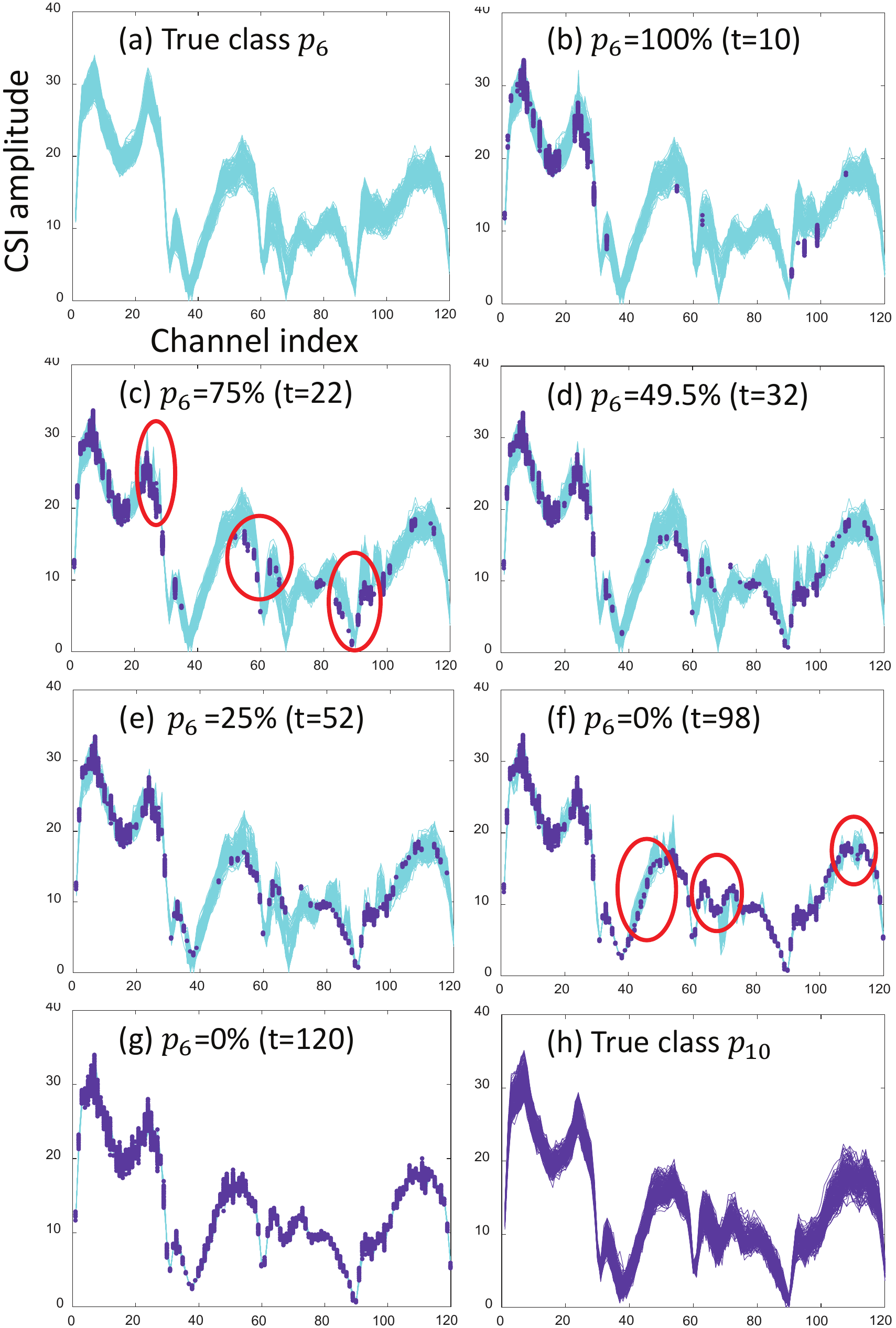}
\caption{$200$ testing samples of CSI of location $p_6$ (in (a)) progressively modified toward $200$ testing samples of CSI of location $p_{10}$ (in (h)) according to the channel ordering sequence $\mathcal{O}_2(p_6\to p_{10})$. The unmodified/modified channels are color-coded by the CSI of $p_6$/$p_{10}$ (in (b)--(g)). The percentages of correct classification as $p_6$ and the numbers of channels modified ($t$) are shown.}
\label{fig:LRP_waveform}
\end{center}
\end{figure}

As will be illustrated (in Figs.~\ref{fig:LRP_waveform}(a) and \ref{fig:LRP_waveform}(h)), the CSI patterns of locations $p_6$ and $p_{10}$ are somewhat similar (by visual inspection) and these two locations are easily misclassified as each other. The raw training samples of locations $p_6$ and $p_{10}$ are close to each other as seen in Fig.~\ref{fig:tsne_training}(a). The recall for location $p_{10}$ for $k$-NN based on the raw CSI data is $96\%$ from Table~\ref{tab:exp1_Precision and Recall}, and the $4\%$ misses have all been classified as location $p_6$. This may explain the fact that when channels of location $p_6$ are modified toward those of location $p_{10}$, all misclassified samples are classified as location $p_{10}$ (i.e., the two topmost curves in Fig.~\ref{fig:LRP_p6_p10_p11} exhibit symmetric trends). For a different pair of locations, e.g., locations $p_{11}$ and $p_{10}$, as compared in Fig.~\ref{fig:LRP_p6_p10_p11}, we observe that the decrease in classification as the true class ($p_{11}$) does not lead to the increase in classification as another class ($p_{10}$) in a symmetric way. Eventually, though, as the majority of channels are such modified, the percentage of classification as the true class decreases to $0\%$ and the percentage of classification as another class increases to $100\%$, regardless of the pairs of locations.

\begin{figure}[t]
\begin{center}
\subfigure[]{
    \label{fig:LRP_heatmap_p6_p6}
    \includegraphics[width=0.47\columnwidth,height=1.13in]{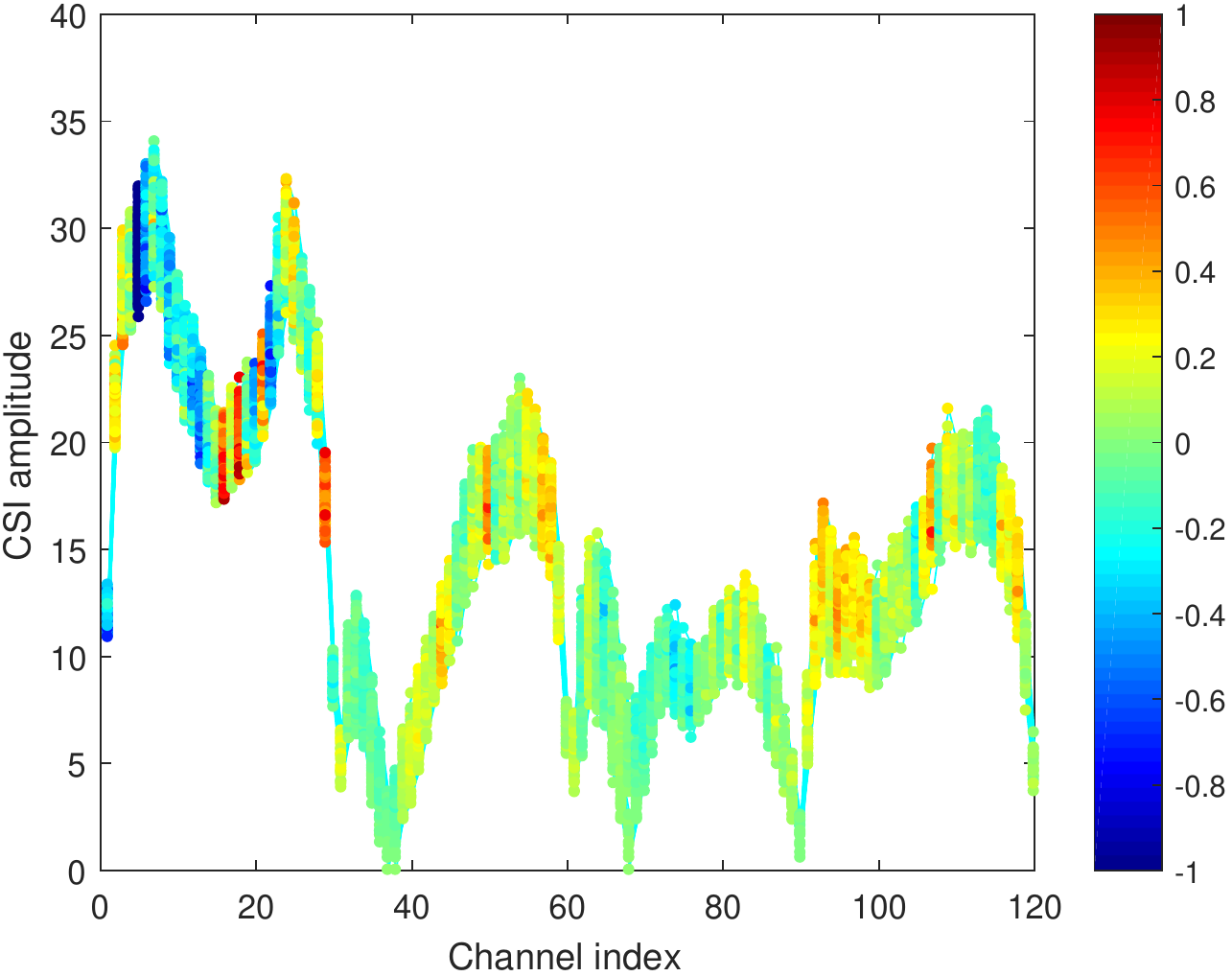}}
\hspace*{-0.06in}
\subfigure[]{
    \label{fig:LRP_heatmap_p6_p10}
    \includegraphics[width=0.47\columnwidth,height=1.13in]{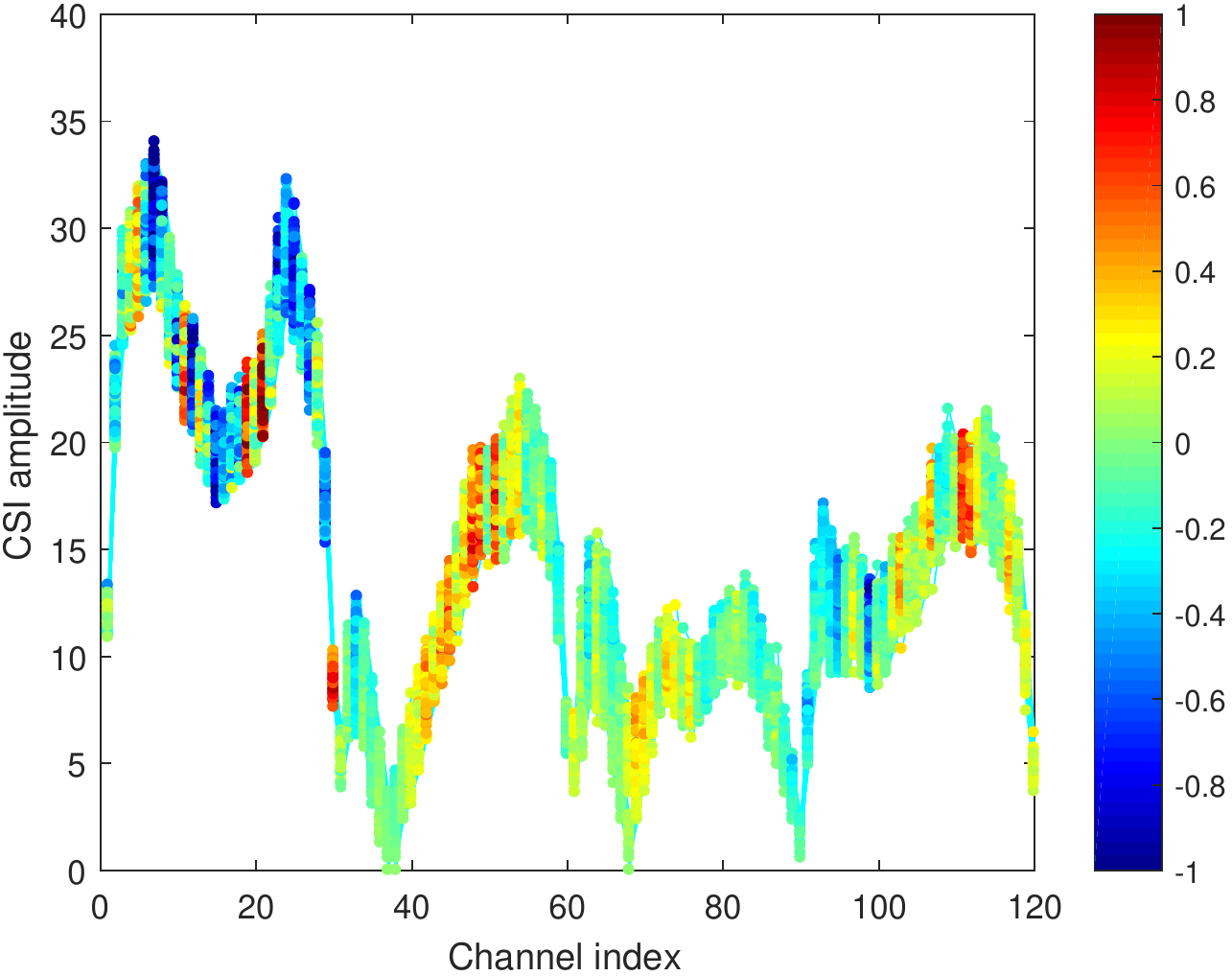}}
\caption{Heatmaps of relevance scores (a) $h'_i(p_6\to p_6)$ and (b) $h'_i(p_6\to p_{10})$ for $i=1,\ldots,120$, superimposed on the $200$ testing samples of CSI of location $p_6$.}
\label{fig:LRP_heatmap}
\end{center}
\end{figure}

Figs.~\ref{fig:LRP_heatmap} and \ref{fig:LRP_waveform} examine the specific channels progressively modified which lead to the result shown by the topmost curve in Fig.~\ref{fig:LRP_p6_p10_p11}. The heatmaps of relevance scores $h'_i(p_6\to p_6)$ and $h'_i(p_6\to p_{10})$ are shown on top of the CSI of location $p_6$ in Fig.~\ref{fig:LRP_heatmap}. Progressive channel modification is visualized in Fig.~\ref{fig:LRP_waveform}. The channel modification starts with the channels with the most negative valued $h'_i(p_6\to p_{10})$, i.e., the blue-colored channels in Fig.~\ref{fig:LRP_heatmap_p6_p10}. When up to $10$ channels are modified, the percentage of correct classification remains $100\%$. When more channels are modified, the percentage starts to decline. Comparing Figs.~\ref{fig:LRP_waveform}(b) and \ref{fig:LRP_waveform}(c), the additional channels modified, roughly corresponding to the red-circled ones in Fig.~\ref{fig:LRP_waveform}(c), are evidence in support of predicting $p_6$ (yellow/red in Fig.~\ref{fig:LRP_heatmap_p6_p6}) but neutral or against predicting $p_{10}$ (cyan/blue in Fig.~\ref{fig:LRP_heatmap_p6_p10}). In Fig.~\ref{fig:LRP_waveform}(c), for the $25\%$ misclassified samples, the channels with $h'_i(p_6\to p_6)>0.4$ (i.e., strong evidence in support of predicting $p_6$) have all been modified. By modifying these critical channels in support of predicting $p_6$, the percentage of correct classification as $p_6$ decreases rapidly (from $100\%$ to about $50\%$ when only $22$ channels are modified, from Fig.~\ref{fig:LRP_waveform}(b) to Fig.~\ref{fig:LRP_waveform}(d)). As channel modification progresses, channels with near-zero and positive valued $h'_i(p_6\to p_{10})$, i.e., the cyan/yellow/red-colored channels in Fig.~\ref{fig:LRP_heatmap_p6_p10}, start to be modified. Comparing Figs.~\ref{fig:LRP_waveform}(e) and \ref{fig:LRP_waveform}(f), the additional channels modified, roughly corresponding to the red-circled ones in Fig.~\ref{fig:LRP_waveform}(f), are evidence neutral or mildly supporting predicting $p_6$ (cyan/yellow in Fig.~\ref{fig:LRP_heatmap_p6_p6}) and supporting predicting $p_{10}$ (yellow/red in Fig.~\ref{fig:LRP_heatmap_p6_p10}). Modifying these channels not critically supporting predicting $p_6$ and against predicting $p_{10}$ results in a slower decrease in the percentage of correct classification (from about $50\%$ to $0\%$ when as many as $66$ channels are modified, from Fig.~\ref{fig:LRP_waveform}(d) to Fig.~\ref{fig:LRP_waveform}(f)). When all $120$ channels of location $p_6$ are modified (Fig.~\ref{fig:LRP_waveform}(g)), the resulting CSI pattern approaches that of location $p_{10}$ (Fig.~\ref{fig:LRP_waveform}(h)).

\subsection{Subcarrier Importance} \label{sec:results_exp1_5}

\begin{figure}[t]
\begin{center}
\includegraphics[width=1\columnwidth]{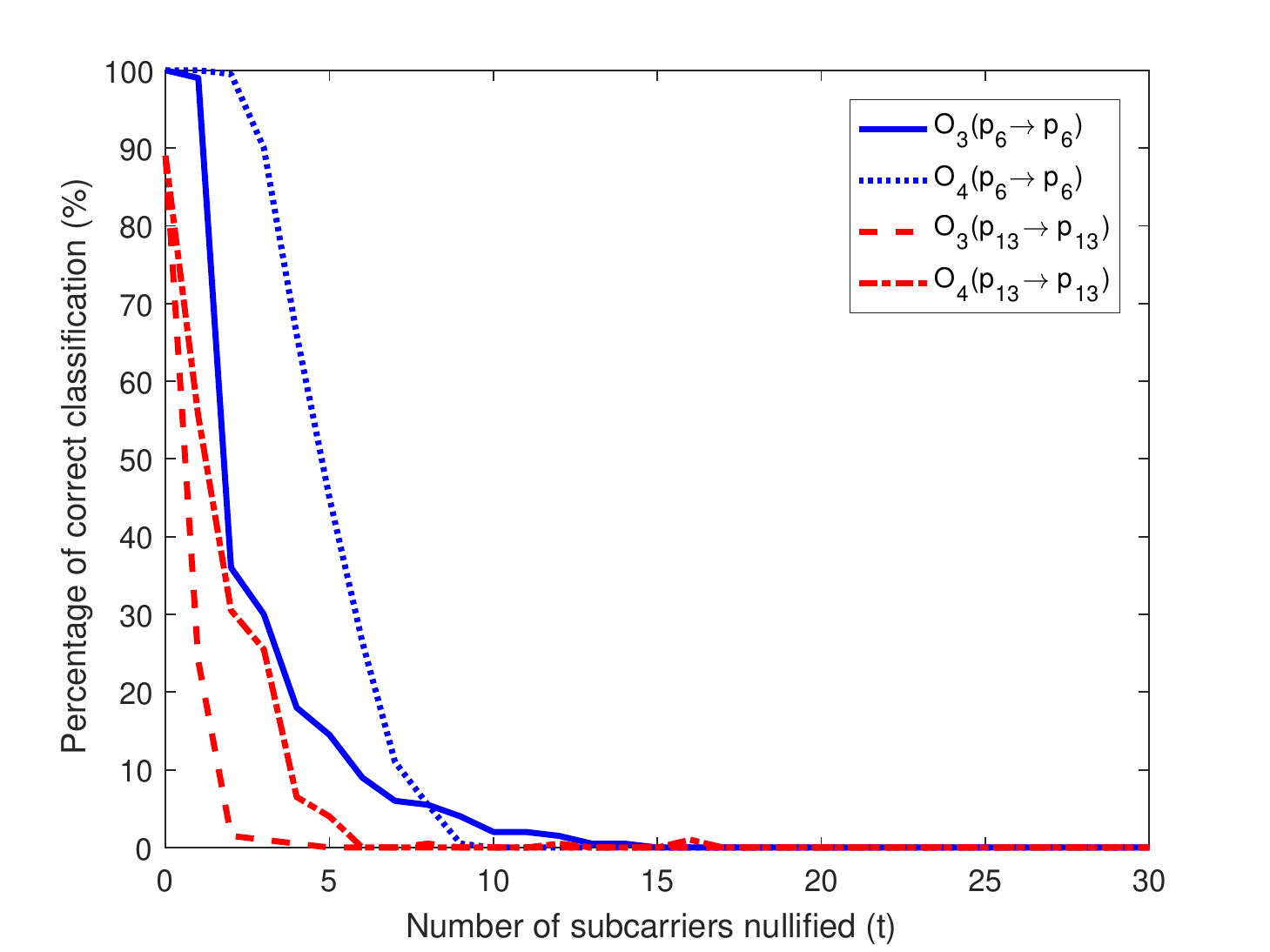}
\caption{The percentage of correct classification (true class: $p_6$ (or $p_{13}$)) after progressive subcarrier nullification according to the subcarrier ordering sequences $\mathcal{O}_3(p_6\to p_6)$ and $\mathcal{O}_4(p_6\to p_6)$ (or $\mathcal{O}_3(p_{13}\to p_{13})$ and $\mathcal{O}_4(p_{13}\to p_{13})$).}
\label{fig:LRP_p6_p13_subcarrier}
\end{center}
\end{figure}

\begin{figure}[t]
\begin{center}
\includegraphics[width=1\columnwidth]{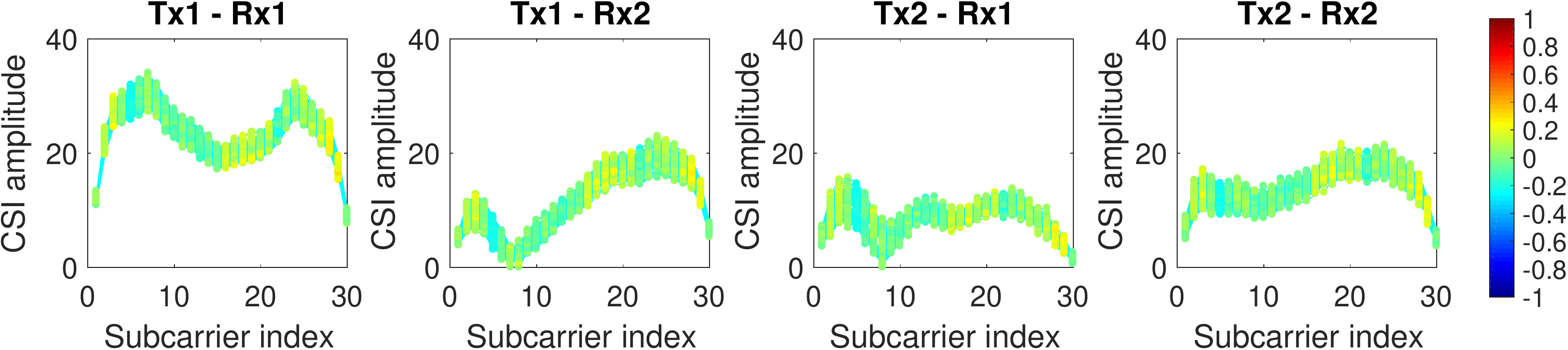}
\caption{Heatmaps of relevance scores $s'_i(p_6\to p_6)$ for $i=1,\ldots,30$, superimposed on the $200$ testing samples of CSI for four transmit-receive antenna pairings of location $p_6$.}
\label{fig:LRP_heatmap_p6_p6_subcarrier}
\end{center}
\end{figure}

From a wireless perspective, the logical $120$ channels comprise $30$ physical subcarriers in $2\times 2$ MIMO wireless links. By breaking down the $120$ channels in view of wireless links and by defining the relevance scores of subcarriers and subcarrier ordering sequences, the subcarrier importance can be examined. Fig.~\ref{fig:LRP_p6_p13_subcarrier} plots the results of progressive subcarrier nullification according to the subcarrier ordering sequences $\mathcal{O}_3$ and $\mathcal{O}_4$ on the DNN prediction for the same two locations as in Fig.~\ref{fig:LRP_p6_p13}. Similar conclusions can be drawn here. Fig.~\ref{fig:LRP_heatmap_p6_p6_subcarrier} shows the heatmaps of subcarrier relevance scores $s'_i(p_6\to p_6)$ on top of the CSI of $2\times 2$ MIMO links for location $p_6$. Two observations can be made. First, the CSI amplitude for a specific subcarrier varies greatly in $2\times 2$ MIMO links, because the four links are independent of each other. Second, the subcarrier relevance scores are more temperate (no extreme values near $1$ or $-1$) than the channel relevance scores in Fig.~\ref{fig:LRP_heatmap_p6_p6}. This suggests that the logical channels, i.e., the $i$th, $(i+30)$th, $(i+60)$th, and $(i+90)$th CSI channels, which all correspond to the same physical subcarrier $i$, do not have relevance scores that are simultaneously high or low (otherwise, it would have led to extreme values after the averaging operation in \eqref{eq:subcarrier_relevance_score}). In other words, the same physical subcarriers at different MIMO links provide distinct information to the DNN prediction. The DNN treats the $120$ logical channels as individual features irrespective of their wireless origins or implications.

\section{Analysis and Visualization of Experimental Results: Experiment 2 (Lounge)} \label{sec:results_exp2}

\begin{figure*}[t]
\begin{center}
\includegraphics[width=2\columnwidth]{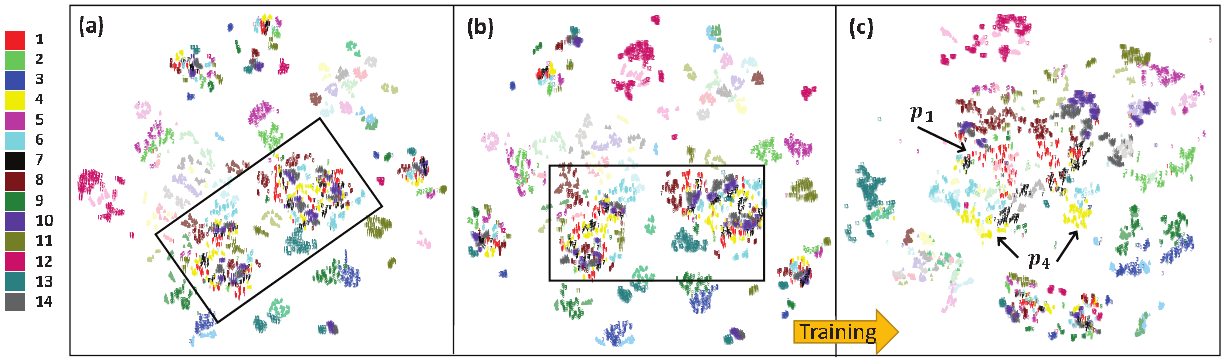}
\caption{The 2D visualization of (a) the raw CSI samples, (b) the last DNN hidden layer activations before training (with random initializations), and (c) the last DNN hidden layer activations after training. For each location, the training samples are shown with a darker shade to distinguish from the testing samples with a lighter shade of the same color. The silhouette scores (calculated for the training samples only) for (a)--(c) are $-0.18$, $-0.21$, and $0.09$, respectively. The rectangular boxes in (a) and (b) enclose all the training samples for locations $p_1$ and $p_4$. Black location markers refer to training samples and colored location markers (in colors corresponding to respective locations) refer to testing samples collected at the respective locations.}
\label{fig:exp2_tsne_training}
\end{center}
\end{figure*}

\begin{figure*}[t]
\begin{center}
\includegraphics[width=2\columnwidth]{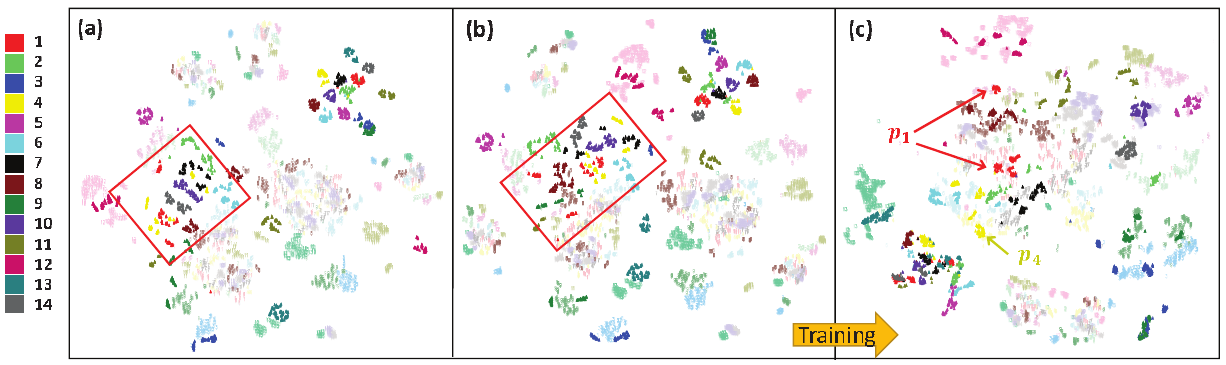}
\caption{Same plots as Fig.~\ref{fig:exp2_tsne_training}, but here, for each location, the testing samples are shown with a darker shade to distinguish from the training samples with a lighter shade of the same color. The silhouette scores (calculated for the testing samples only) for (a)--(c) are $-0.22$, $-0.23$, and $-0.15$, respectively. The rectangular boxes in (a) and (b) enclose the testing samples for locations $p_1$ and $p_4$. Black location markers refer to training samples and colored location markers (in colors corresponding to respective locations) refer to testing samples collected at the respective locations.}
\label{fig:exp2_tsne_testing}
\end{center}
\end{figure*}

The dataset collected in the lounge experiment is more challenging for classification than the dataset collected in the conference room experiment. The lounge environment has more obstructions that produce more disturbance to the wireless signals and drastically different multipath profiles. Specifically, the CSI is more noisy/dispersive across different samples, and each sample exhibits more strongly varying CSI across different subcarriers as a result of richer multipath effects in this environment.

\subsection{Training Effects} \label{sec:results_exp2_1}

The 2D visualization of raw CSI samples shows poor visual separation between classes in Fig.~\ref{fig:exp2_tsne_training}(a), with a silhouette score of $-0.18$. Comparing Figs.~\ref{fig:exp2_tsne_training}(b) and \ref{fig:exp2_tsne_training}(c), the trained DNN yields improved visual separation than the untrained DNN (silhouette scores $-0.21$ vs. $0.09$). The improvement, albeit much smaller as compared to the counterpart in the conference room experiment, suggests that DNN still manages to learn distinctive patterns in this more noisy environment, which leads to better classification results in the testing. Consider locations $p_1$ and $p_4$ as an example, as marked in Figs.~\ref{fig:exp2_tsne_training}(a)--(c). It is seen that before training the samples of the same location are more scattered (as enclosed by the rectangular box) while after training they become better clustered. This leads to reduced misclassification as locations $p_1$ and $p_4$ in the testing, as reflected by a higher precision in the DNN scheme as compared to the $k$-NN scheme for locations $p_1$ and $p_4$, as shown in Table~\ref{tab:exp2_Precision and Recall}. Overall, DNN yields a higher macro-average precision as compared to $k$-NN ($73.29\%$ vs. $62.29\%$).

\subsection{Testing Performance} \label{sec:results_exp2_2}

The poorer visual separation for the raw testing samples, as shown in Fig.~\ref{fig:exp2_tsne_testing}(a), results in a much lower macro-average recall for $k$-NN in the lounge experiment as compared to in the conference room experiment ($47.5\%$ vs. $70.88\%$). After DNN training, it is observed that a better clustering is achieved in Fig.~\ref{fig:exp2_tsne_testing}(c). For example, the raw testing samples for locations $p_1$ and $p_4$, as enclosed by the rectangular box in Fig.~\ref{fig:exp2_tsne_testing}(a), are highly scattered, resulting in a high misclassification rate of the testing samples of location $p_1$ ($p_4$) to other locations. In comparison, after DNN training, as shown in Fig.~\ref{fig:exp2_tsne_testing}(c), the testing samples are better clustered and closer to the training samples of the same class. As a result, DNN yields a higher recall than $k$-NN for locations $p_1$ and $p_4$ from Table~\ref{tab:exp2_Precision and Recall}. The macro-average recall for DNN is higher than for $k$-NN ($63.71\%$ vs. $47.5\%$ in Table~\ref{tab:exp2_Precision and Recall}).

The improvement in silhouette scores between untrained and trained DNN at the last DNN hidden layer (Figs.~\ref{fig:exp2_tsne_testing}(b) and \ref{fig:exp2_tsne_testing}(c)) is smaller in the lounge experiment than in the conference room experiment. This is because there are still testing samples that cannot be well-clustered by DNN (e.g., those located in the lower left corner in Fig.~\ref{fig:exp2_tsne_testing}(c)) in this more noisy environment. However, DNN training still has significant contribution to the classification accuracy, where the difference in the macro-average recall between $k$-NN and DNN is greater in the lounge experiment than in the conference room experiment ($16.21\%$ vs. $8.49\%$).

\subsection{Discriminative Features} \label{sec:results_exp2_3}

\begin{figure}[t]
\begin{center}
\includegraphics[width=1\columnwidth]{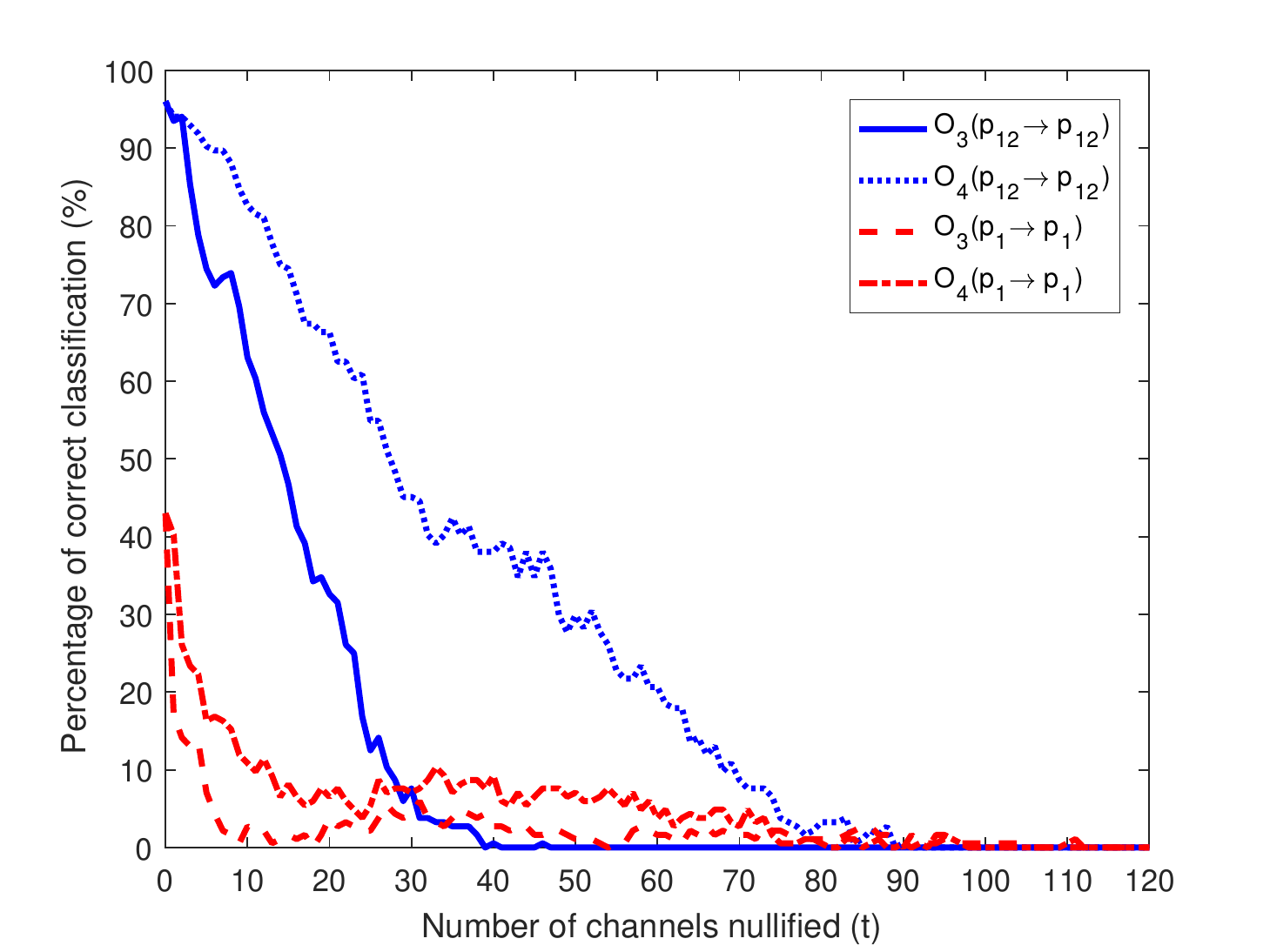}
\caption{The percentage of correct classification (true class: $p_{12}$ (or $p_1$)) after progressive channel nullification according to the channel ordering sequences $\mathcal{O}_3(p_{12}\to p_{12})$ and $\mathcal{O}_4(p_{12}\to p_{12})$ (or $\mathcal{O}_3(p_1\to p_1)$ and $\mathcal{O}_4(p_1\to p_1)$).}
\label{fig:exp2_LRP_p12_p1}
\end{center}
\end{figure}

\begin{figure}[t]
\begin{center}
\includegraphics[width=1\columnwidth]{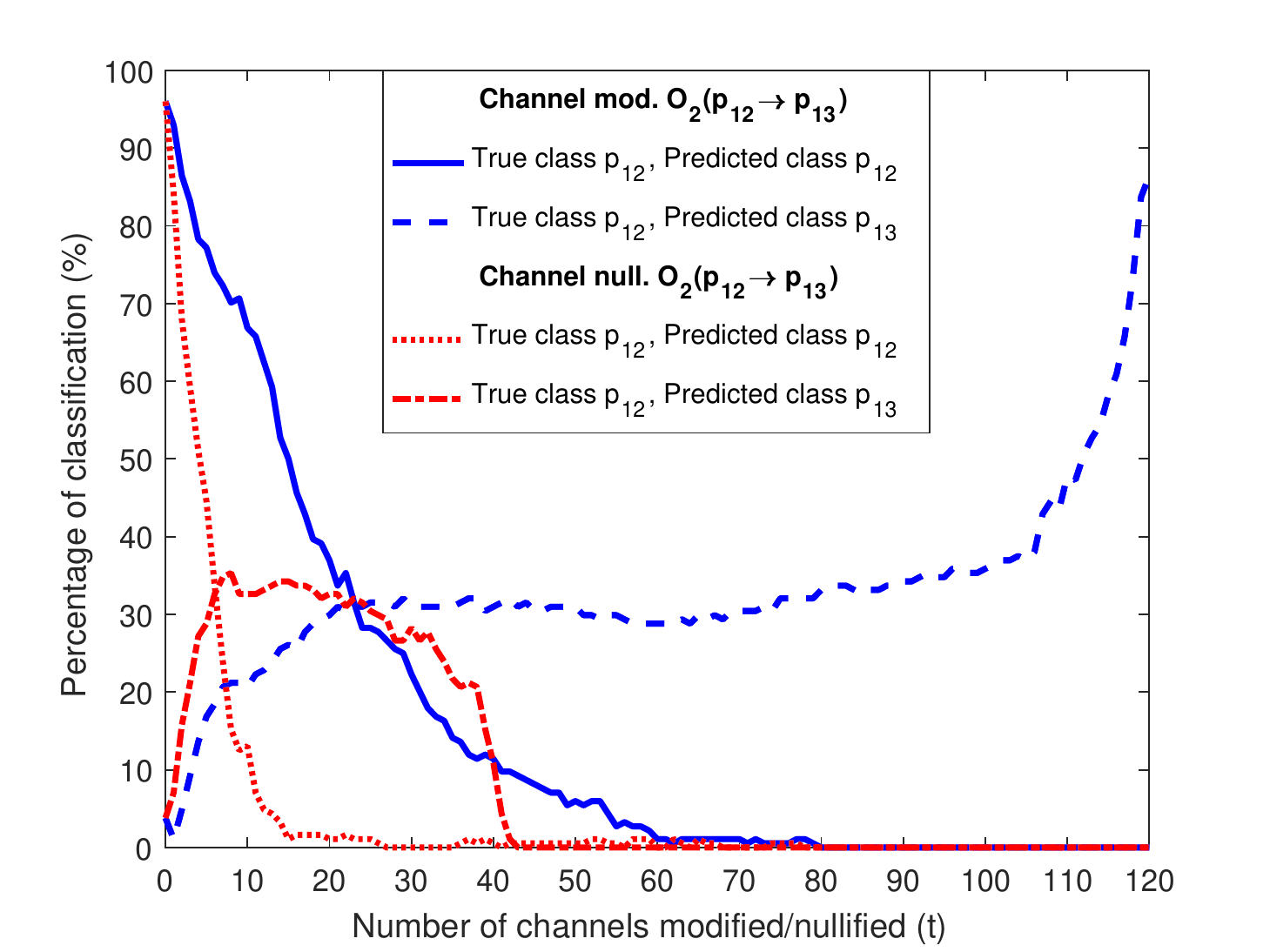}
\caption{The percentage of classification as $p_{12}$ and $p_{13}$ (true class: $p_{12}$) after progressive channel modification according to the channel ordering sequence $\mathcal{O}_2(p_{12}\to p_{13})$, in comparison with the same setting but after progressive channel nullification.}
\label{fig:exp2_LRP_p12_p13}
\end{center}
\end{figure}

\begin{figure}[t]
\begin{center}
\subfigure[]{
    \label{fig:exp2_LRP_heatmap_p12_p12}
    \includegraphics[width=0.47\columnwidth,height=1.13in]{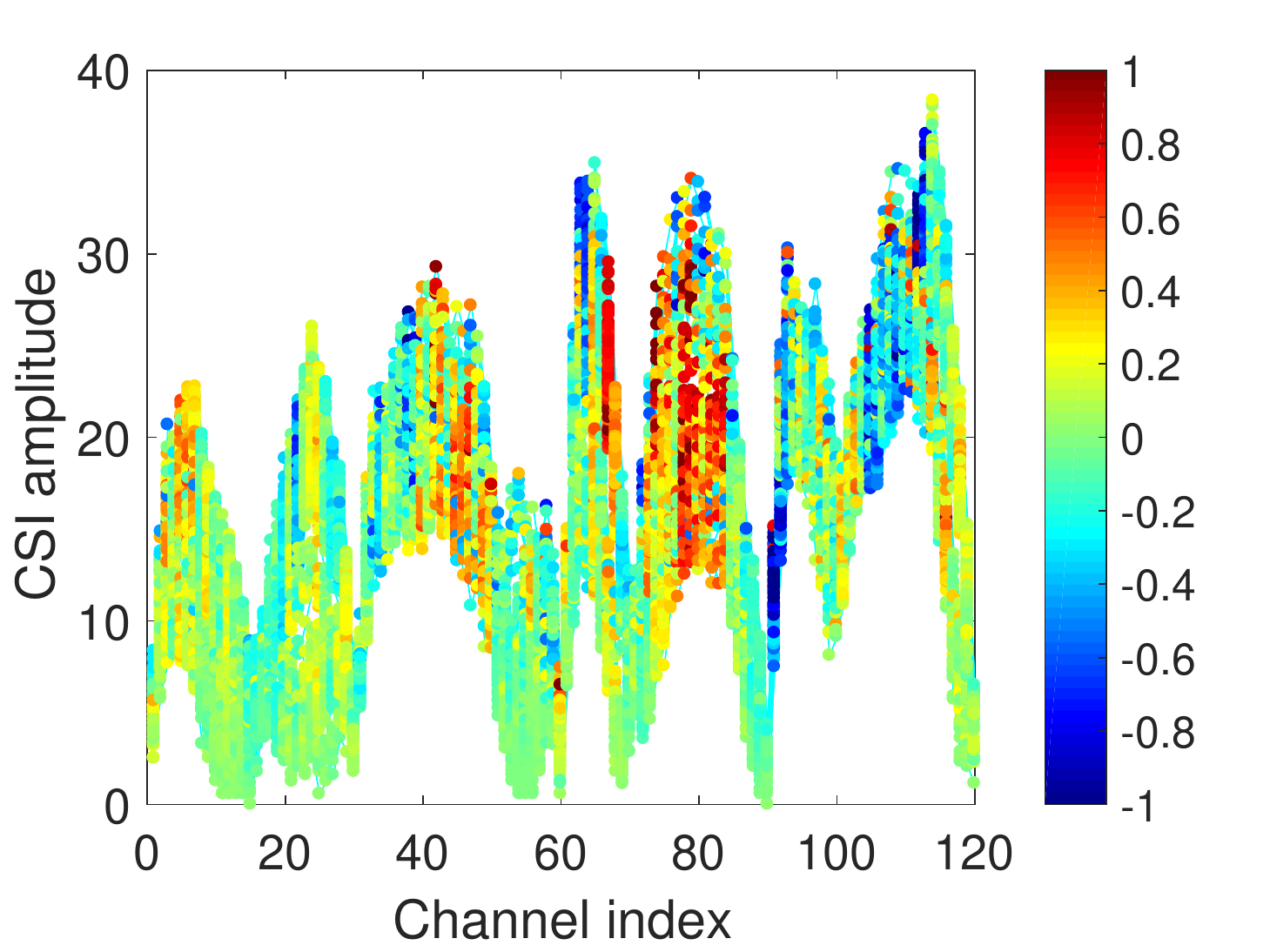}}
\hspace*{-0.06in}
\subfigure[]{
    \label{fig:exp2_LRP_heatmap_p12_p13}
    \includegraphics[width=0.47\columnwidth,height=1.13in]{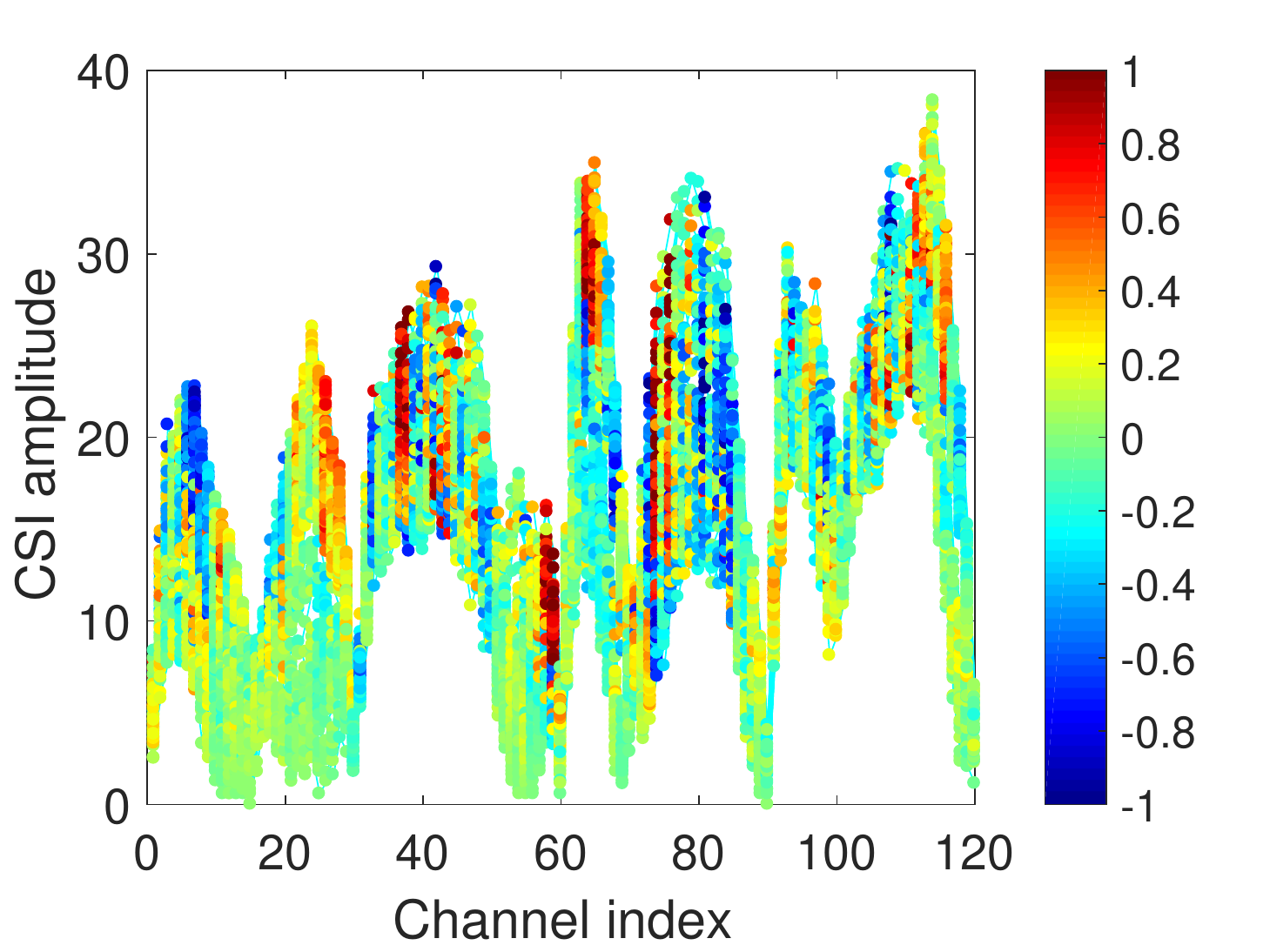}}
\caption{Heatmaps of relevance scores (a) $h'_i(p_{12}\to p_{12})$ and (b) $h'_i(p_{12}\to p_{13})$ for $i=1,\ldots,120$, superimposed on the $184$ testing samples of CSI of location $p_{12}$.}
\label{fig:exp2_LRP_heatmap}
\end{center}
\end{figure}

\begin{figure}[t]
\begin{center}
\includegraphics[width=\columnwidth]{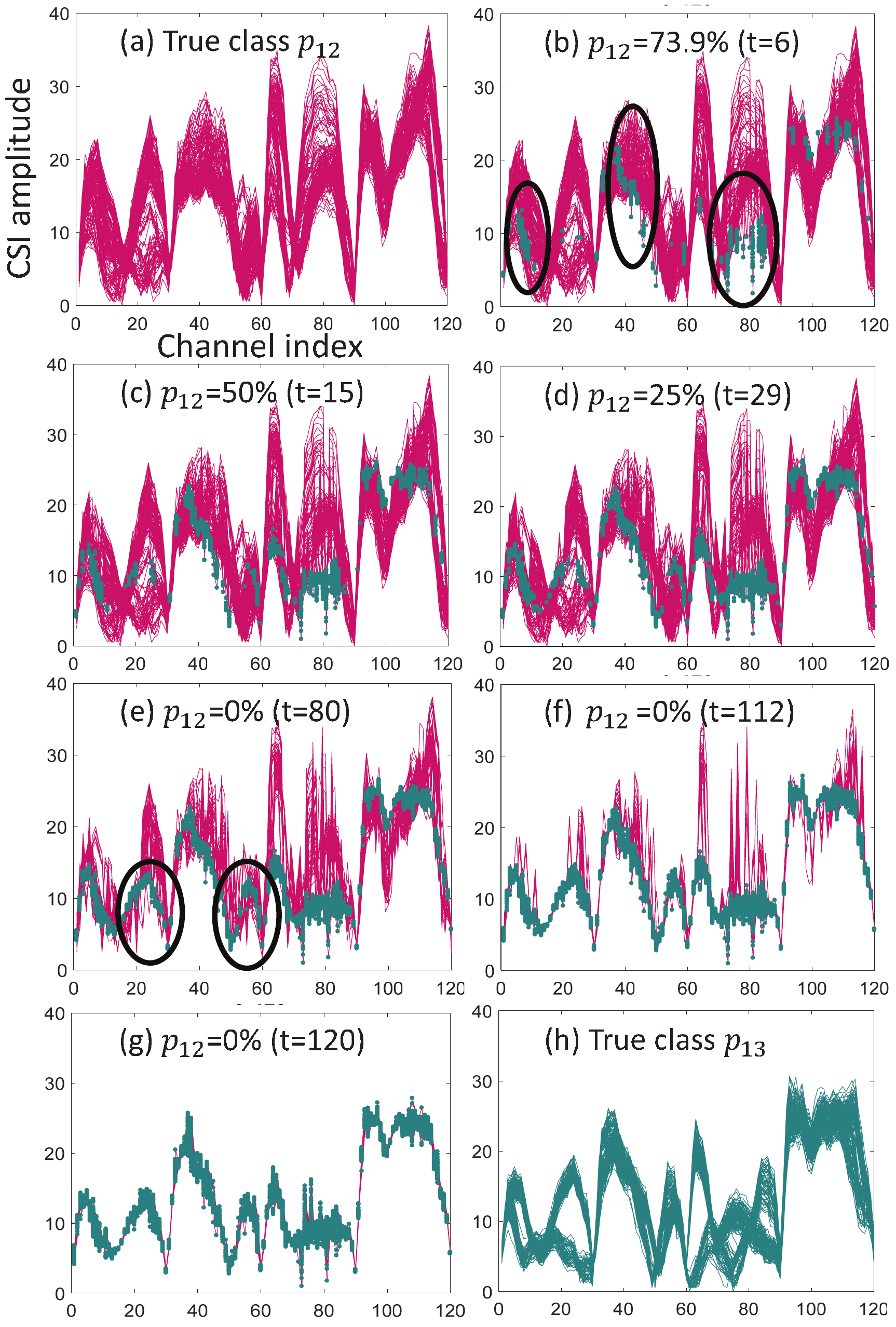}
\caption{$184$ testing samples of CSI of location $p_{12}$ (in (a)) progressively modified toward $184$ testing samples of CSI of location $p_{13}$ (in (h)) according to the channel ordering sequence $\mathcal{O}_2(p_{12}\to p_{13})$. The unmodified/modified channels are color-coded by the CSI of $p_{12}$/$p_{13}$ (in (b)--(g)). The percentages of correct classification as $p_{12}$ and the numbers of channels modified ($t$) are shown.}
\label{fig:exp2_LRP_waveform}
\end{center}
\end{figure}

Similar to the conference room experiment, we show the results of progressive channel nullification for a location near the LoS ($p_{12}$) with a high recall ($96\%$) and a location away from the LoS ($p_1$) with a low recall ($43\%$), in Fig.~\ref{fig:exp2_LRP_p12_p1}. The general observations agree with those in the conference room experiment.

Fig.~\ref{fig:exp2_LRP_p12_p13} shows the effect of progressive channel modification from $p_{12}$ toward $p_{13}$. It is observed that, again, the decrease in classification as the true class ($p_{12}$) generally leads to the increase in classification as $p_{13}$. However, even when all channels are such modified, the percentage of classification as $p_{13}$ does not reach $100\%$. This is because the CSI samples are apparently more noisy/dispersive in this lounge experiment than in the conference room experiment (the waveforms of different CSI samples collectively appear ``thicker'' in Fig.~\ref{fig:exp2_LRP_waveform} than in Fig.~\ref{fig:LRP_waveform}), and thus there is a higher discrepancy between the modified CSI patterns toward $p_{13}$ and the actual CSI patterns of $p_{13}$ (comparing Figs.~\ref{fig:exp2_LRP_waveform}(g) and \ref{fig:exp2_LRP_waveform}(h)). The cross-comparison between channel modification and nullification concludes similarly as in the conference room experiment.

Figs.~\ref{fig:exp2_LRP_heatmap} and \ref{fig:exp2_LRP_waveform} examine the specific channels progressively modified which lead to the results in Fig.~\ref{fig:exp2_LRP_p12_p13}. Channel modification is more sensitive in this lounge experiment, as the percentage of predicting $p_{12}$ declines rapidly as channels are modified. By modifying only six channels for each sample, the percentage of classifying as $p_{12}$ decreases to $73.9\%$ while the percentage of classifying as $p_{13}$ increases to $18.48\%$. These channels initially modified, roughly corresponding to the black-circled ones in Fig.~\ref{fig:exp2_LRP_waveform}(b), represent evidence in support of predicting $p_{12}$ (yellow/red in Fig.~\ref{fig:exp2_LRP_heatmap_p12_p12}) but neutral or against predicting $p_{13}$ (cyan/blue in Fig.~\ref{fig:exp2_LRP_heatmap_p12_p13}). Comparing Figs.~\ref{fig:exp2_LRP_waveform}(d) and \ref{fig:exp2_LRP_waveform}(e), the additional $51$ channels modified, roughly corresponding to the black-circled ones in Fig.~\ref{fig:exp2_LRP_waveform}(e), are evidence neutral for predicting both $p_{12}$ and $p_{13}$, leading to a slow decrease and increase in the percentages of predicting $p_{12}$ and $p_{13}$, respectively. Comparing Figs.~\ref{fig:exp2_LRP_waveform}(g) and \ref{fig:exp2_LRP_waveform}(h), there is a more pronounced discrepancy between the modified CSI toward the mean of CSI of $p_{13}$ and the actual CSI of $p_{13}$, due to a larger variation in the testing CSI samples in the lounge experiment.

\subsection{Subcarrier Importance} \label{sec:results_exp2_4}

\begin{figure}[t]
\begin{center}
\includegraphics[width=1\columnwidth]{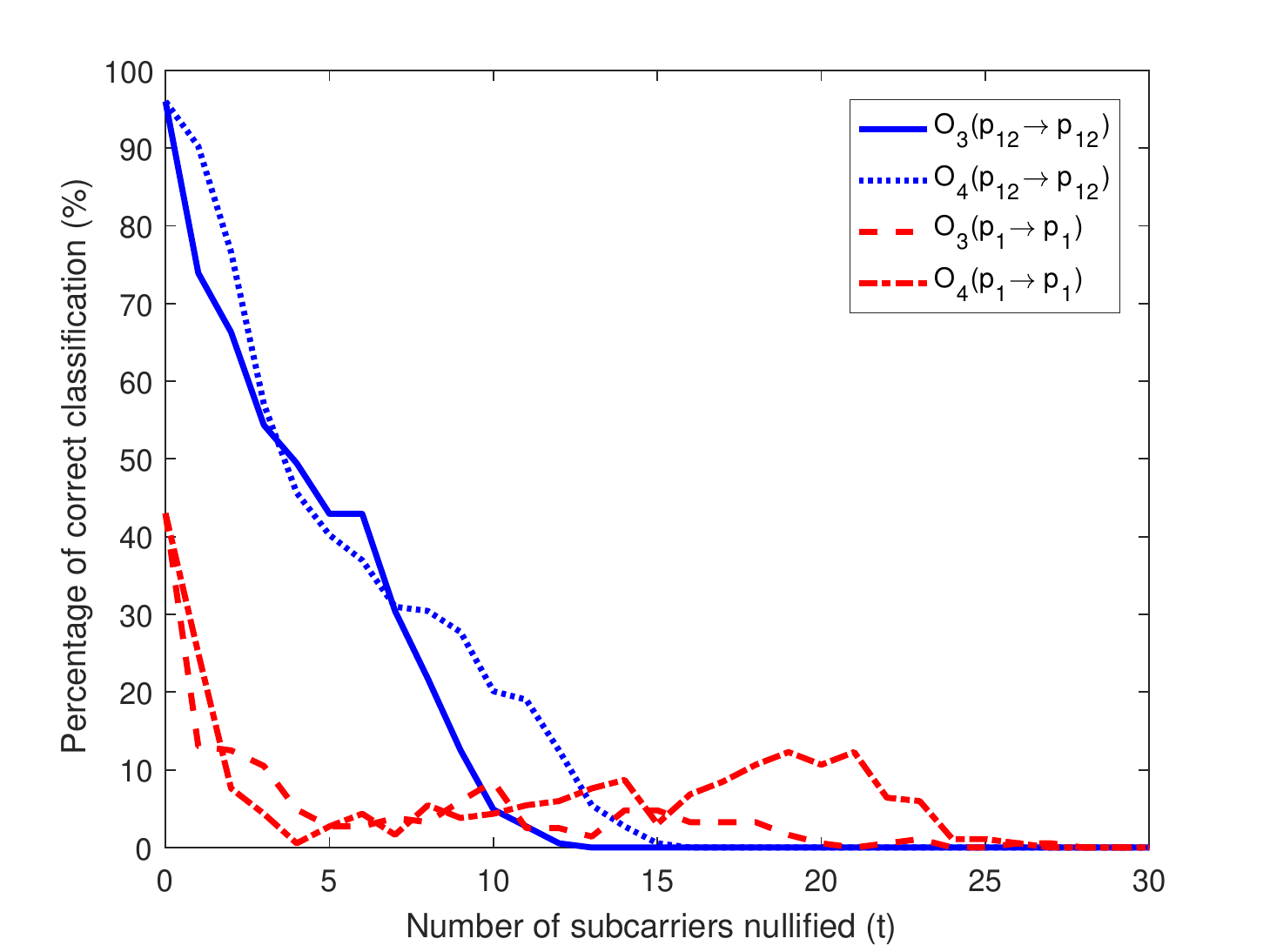}
\caption{The percentage of correct classification (true class: $p_{12}$ (or $p_1$)) after progressive subcarrier nullification according to the subcarrier ordering sequences $\mathcal{O}_3(p_{12}\to p_{12})$ and $\mathcal{O}_4(p_{12}\to p_{12})$ (or $\mathcal{O}_3(p_1\to p_1)$ and $\mathcal{O}_4(p_1\to p_1)$).}
\label{fig:exp2_LRP_p12_p1_subcarrier}
\end{center}
\end{figure}

\begin{figure}[t]
\begin{center}
\includegraphics[width=1\columnwidth]{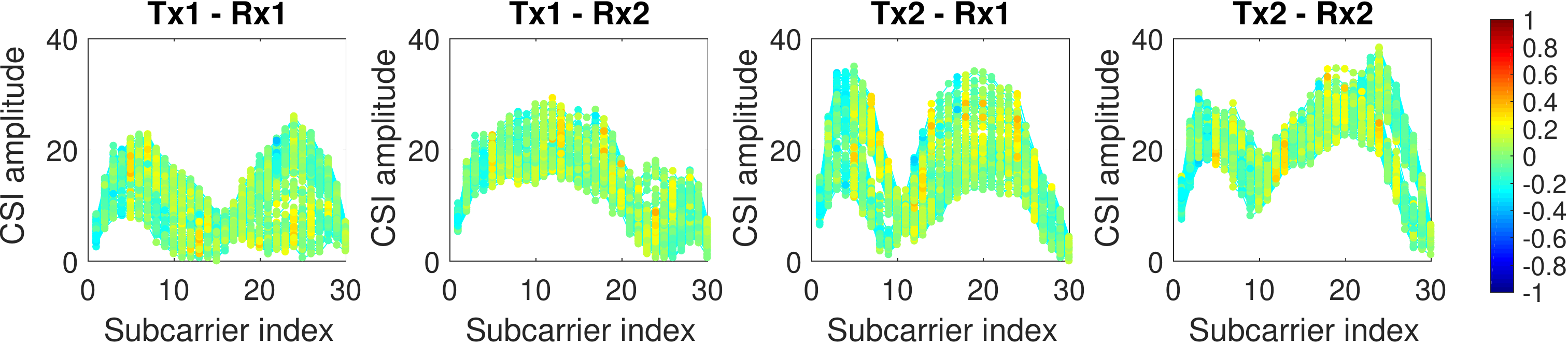}
\caption{Heatmaps of relevance scores $s'_i(p_{12}\to p_{12})$ for $i=1,\ldots,30$, superimposed on the $184$ testing samples of CSI for four transmit-receive antenna pairings of location $p_{12}$.}
\label{fig:exp2_LRP_heatmap_p12_p12_subcarrier}
\end{center}
\end{figure}

Figs.~\ref{fig:exp2_LRP_p12_p1_subcarrier} and \ref{fig:exp2_LRP_heatmap_p12_p12_subcarrier} present results related to subcarrier importance. There is a richer multipath effect (larger delay spread) in the lounge environment than in the conference room environment, as shown by the generally smaller correlation coefficients between two adjacent subcarriers in the lounge experiment (comparing the same pairs of adjacent subcarriers, the lounge environment yields smaller correlation coefficients in $26$ out of the total $29$ pairs of adjacent subcarriers). In the lounge experiment, the logical channels that correspond to the same physical subcarriers have even more greatly varying relevance scores. Thus, progressively nullifying a specific subcarrier in four wireless links simultaneously translates to channel nullification in neither the channel ordering sequence of $\mathcal{O}_3$ nor $\mathcal{O}_4$. Thus, Fig.~\ref{fig:exp2_LRP_p12_p1_subcarrier} exhibits a slightly different trend as compared to Fig.~\ref{fig:exp2_LRP_p12_p1}. The richer multipath effect, which, in theory, provides more degrees of freedom which could aid feature selection for classification, may have led to a greater performance improvement of DNN over $k$-NN in the macro-average recall and precision in the lounge experiment (see Tables~\ref{tab:exp1_Precision and Recall} and \ref{tab:exp2_Precision and Recall}). Yet, in the meantime, the same environment induces more noisy CSI across different samples, which deteriorates the performance of DNN (as well as $k$-NN and SVM) in absolute terms in the lounge experiment.

\section{Conclusion} \label{sec:conclusion}

In this paper, we have performed a detailed analysis of the DNN in device-free Wi-Fi fingerprinting indoor localization through visual analytics techniques. By projecting the last DNN hidden layer activations to visualizable 2D representations via the t-SNE dimensionality reduction technique, we observed better clustering and visual separation among different classes after DNN training. The more concentrated distribution of the testing samples matches the reduced misclassification in the testing for DNN. DNN generally yields higher precision results as compared to $k$-NN based on the raw data, suggesting that DNN training is effective in learning discriminative features from the noisy wireless signals. The LRP visual analytics technique, coupled with the channel nullification/modification procedures, revealed critical channels or features in wireless signals learned by the DNN which cannot be well interpreted by human perceptions. The relevance scores, determined by LRP, provided analyzable, interpretable, and reliable measures of how and why the percentages of classification change as specific channels or features are manipulated, and thereby allow for a better understanding of the mechanism of DNN-based device-free wireless indoor localization.

\bibliographystyle{IEEEtran}
\bibliography{IEEEabrv,ref}
\balance

\end{document}